\documentclass{aa}  

\usepackage{float}
\usepackage{graphicx}
\usepackage{txfonts}
\usepackage{natbib}
\bibpunct{(}{)}{,}{a}{}{,}  
\usepackage{pdflscape}
\usepackage{subcaption}
\usepackage{placeins}
\usepackage{hyperref}
\hypersetup{breaklinks=true,
            colorlinks,
            filecolor=blue,
            linkcolor=blue,
            citecolor=blue,
            urlcolor=blue,
            anchorcolor=blue} 
\usepackage{scalerel}

\begin{document}

   \title{The {\it Hubble} Missing Globular Cluster Survey }

   \subtitle{I. Survey overview and the first precise age estimate for ESO452-11 and 2MASS-GC01}
   
   \titlerunning{{\it Hubble} MGCS. I}
   
   \author{D. Massari\inst{1}
          \and
          M. Bellazzini\inst{1}
          \and
          M. Libralato\inst{2}
          \and
          A. Bellini\inst{3}
          \and
          E. Dalessandro\inst{1}
          \and
          E. Ceccarelli\inst{1,4}
          \and
          F. Aguado-Agelet\inst{5,6}
          \and
          S. Cassisi\inst{7,8}
          \and
          C. Gallart\inst{6,9}
          \and
          M. Monelli\inst{6,9,10}
          \and
          A. Mucciarelli\inst{4,1}
          \and
          E. Pancino\inst{11}
          \and
          M. Salaris\inst{12,7}
          \and
          S. Saracino\inst{11, 12}
          \and 
          E. Dodd\inst{13}
          \and 
          F. R. Ferraro\inst{4}
          \and 
          E. R. Garro\inst{14}
          \and
          B. Lanzoni\inst{4}
          \and
          R. Pascale\inst{1}
          \and
          L. Rosignoli\inst{4, 1}
          }

   \institute{INAF - Astrophysics and Space Science Observatory of 
              Bologna, Via Gobetti 93/3, 40129 Bologna, Italy\\ \email{davide.massari@inaf.it}
         \and
            INAF, Osservatorio Astronomico di Padova, Vicolo dell’Osservatorio 5, Padova, I-35122, Italy
         \and 
            Space Telescope Science Institute, 3700 San Martin Drive, Baltimore, MD 21218, USA
         \and
             Department of Physics and Astronomy, University of 
             Bologna, Via Gobetti 93/2, 40129 Bologna, Italy 
         \and
             atlanTTic, Universidade de Vigo, Escola de Enxeñar\'ia de Telecomunicaci\'on, 36310, Vigo, Spain 
         \and
             Universidad de La Laguna, Avda. Astrof\'isico Fco. S\'anchez, E-38205 La Laguna, Tenerife, Spain
         \and
             INAF – Osservatorio Astronomico di Abruzzo, Via M. Maggini, 64100 Teramo, Italy
         \and
             INFN - Sezione di Pisa, Universit\'a di Pisa, Largo Pontecorvo 3, 56127 Pisa, Italy
         \and
            Instituto de Astrof\'isica de Canarias, Calle V\'ia L\'actea s/n, E-38206 La Laguna, Tenerife, Spain
         \and 
            INAF – Osservatorio Astronomico di Roma, Via Frascati 33, 00078 Monte Porzio Catone, Roma, Italy
         \and
            INAF - Osservatorio Astrofisico di Arcetri, Largo E. Fermi 5, I-50125 Firenze, Italy
         \and
            Astrophysics Research Institute, Liverpool John Moores University, 146 Brownlow Hill, Liverpool L3 5RF, UK     
         \and
             Institute for Computational Cosmology \& Centre for Extragalactic Astronomy, Department of Physics, Durham University, South Road,
             Durham, DH1 3LE, UK
         \and
            ESO – European Southern Observatory, Alonso de Cordova 3107, Vitacura, Santiago, Chile
}   
   \date{Received XX; accepted YY}

% \abstract{}{}{}{}{} 
% 5 {} token are mandatory
 
  \abstract
 {We present the {\it Hubble} Missing Globular Cluster Survey (MGCS), a {\it Hubble Space Telescope} Treasury Program dedicated to the observation of all kinematically confirmed Milky Way globular clusters that missed previous {\it Hubble} imaging. After introducing the aims of the programme and describing its target clusters, we showcase the first results of the survey. These are related to two clusters, one located at the edge of the Milky Way bulge and observed in optical bands, namely ESO452-11, and one located in the Galactic disc observed in the near-IR, namely 2MASS-GC01. For both clusters, the deep colour-magnitude diagrams obtained from the MGCS observations reach several magnitudes below their main-sequence turn-off and thus enable the first precise estimate of their age. By using the methods developed in the Cluster Ages to Reconstruct the Milky Way Assembly (CARMA) project, we find ESO452-11 to be an old metal-intermediate globular cluster, with ${\rm [M/H]}\simeq-0.80^{+0.08}_{-0.11}$ and an age of ${\rm t}=13.59^{+0.48}_{-0.69}$ Gyr. Its location on the age-metallicity relation makes it consistent with an in situ origin, in agreement with its dynamical properties. On the other hand, the results for 2MASS-GC01 highlight it as a young metal-intermediate cluster, with an age of ${\rm t}=7.22^{+0.93}_{-1.11}$ Gyr at ${\rm [M/H]}=-0.73^{+0.06}_{-0.06}$. Despite the large associated uncertainty, our age estimate for this extremely extincted cluster indicates it to be either the youngest globular cluster known to date or a massive and compact open cluster, which is consistent with its almost circular, disc-like orbit.}


   \maketitle
%
%-------------------------------------------------------------------

\section{Introduction}

Globular clusters (GCs) are among the most powerful tracers of the early history of the Milky Way (MW). Investigating the system of MW GCs with the aim of reconstructing the formation and assembly of the Galactic halo dates back to the seminal works of \citet{kinman59} and \citet{searle78}. Since then, many works have combined the information on the intrinsic properties of GCs, such as age \citep[see e.g.][]{marin-franch09, forbes2010, leaman13, kruijssen19, massari23} or chemistry \citep[e.g.][]{fall85, recio-blanco18, horta20, monty23, ceccarelli24a}, to understand the origin of individual GCs: for example, whether they were born in situ or accreted during past merger events.
%tell 
%the fraction of clusters born {\it in-situ} apart from those accreted during past merger events. 
With the systematic availability of dynamical information on GC orbits, achieved thanks to the advent of the {\it Gaia} mission \citep[][]{gaia}, this effort has made huge steps forward. The addition of this information has in fact led to a rather robust classification of in situ and accreted GCs \citep[][]{massari19, forbes20, callingham22, malhan22, belokurov24, chen24}, and to the first attempts to identify the different galaxy progenitors of the accreted ones.  

The next step forward in investigating MW GCs as tracers of Galactic evolution is improving the accuracy of the measurement of their properties. From the kinematic point of view, this has already been achieved thanks to the homogeneous measurement, by the same instrument \citep[the ESA-{\em Gaia} mission; see e.g.][and references therein]{gaia, gaiadr3}, of the cluster positions and proper motions \citep[see e.g.][]{vasiliev21}. The measurement of GC distance still suffers from inhomogeneity in the adopted methods, although attempts to overcome the related systematic errors have led to typical accuracies of a few per cent \citep{baumgardt21}.
From a chemical point of view, things are more complex. The sources of systematic offsets among different abundance measurements are many and range from the instrumental setups (wavelength range, spectral resolution, etc.) to the assumptions on the atmospheric models and the atomic parameters, and to the methods used to estimate the atmospheric parameters and the elemental abundances. In this case, increasing effort is also being dedicated to improving the abundance homogeneity between different sources \citep[see e.g.][]{thomas24}, but offsets as large as a few 0.1 dex are still common for more than one element \citep[e.g.][]{carretta23, schiavon24}.
 
Concerning age measurements and, more generally, any photometric study, a single facility has historically revolutionised our knowledge of GCs: the {\it Hubble Space Telescope (HST)}.
Its exquisite spatial resolution, coupled with high sensitivity in passbands ranging from the ultraviolet (UV) to the near-infrared (NIR), has paved the way for a monumental advance in a plethora of GC-related science cases. These range from the investigation of GC multiple populations \citep[e.g.][]{milone12}, to the study of their ages \citep{marin-franch09,dotter10}, mass function \citep[e.g.][]{paust10}, and binary fractions \citep[e.g.][]{milone12}, and to the characterisation of their hottest stars such as white dwarfs \citep[e.g.][]{bedin23}, blue and extreme horizontal
branch stars \citep[e.g.][]{brown16}, and blue stragglers \citep[e.g.][]{ferraro12}. Some of these achievements were obtained by means of many focused General Observers (GO) HST programmes, but the advent of large observational campaigns such as the Advanced Camera for Surveys (ACS) Survey of Galactic Globular Clusters \citep{sarajedini07}, or the HST UV Legacy Survey of Galactic Globular Clusters \citep{piotto15}, has effectively shaped the state of the art of GC science; and more than ten years since their completion, these data still have huge value for the community.
In spite of the monumental scientific return of existing HST observations, the heritage of HST for Galactic GCs is still very incomplete. Taking into account the most recent discoveries of new Milky Way GCs, there are still 33 kinematically confirmed GCs, either via proper
motions \citep{vasiliev21} or via radial velocities \citep[see][]{baumgardt21, pace23, garro23}, that lack any kind of HST imaging, plus one that only has observations in one passband.

The Missing Globular Cluster Survey (MGCS; HST Treasury Program GO-17435, PI: Massari) targets these 34 GCs with the aim of providing homogeneous two-band photometry and astrometry for each of them.
The survey has three main objectives:
{\it Age determination}. The main goal of the {\it Hubble} MGCS is to determine the age of the targets.  With the precision enabled by HST imaging and the methods developed by the Clusters Age to Reconstruct the Milky Way Assembly (CARMA) collaboration \citep[$< 500$ Myr on relative ages;][]{vandenberg13, massari23}, MGCS will estimate the age of the 34 clusters, sampling for the first time several magnitudes below their Main Sequence Turn-Off (MSTO). The age estimates will be used in combination with metallicity measurements and the clusters' orbital properties derived from {\it Gaia} data, to robustly assess their former galaxy progenitor \citep[see e.g.][]{massari19, callingham22}, and to derive fundamental properties of the merger events that have led them into the Milky Way, such as merger mass and accretion time \citep[][]{kruijssen19}.\\ 
{\it Search for bulge fossil relics}. Deep, high-resolution imaging of star clusters in the Galactic bulge has revealed the existence of a class of systems, called “bulge fossil fragments”, that might be important contributors to the build-up of the MW bulge \citep[][]{ferraro21}. These systems are all characterised by very complex colour-magnitude diagrams (CMDs) indicating the coexistence of multi-age and multi-metallicity sub-populations, such as in the case of Terzan 5 \citep[][]{ferraro09, ferraro16, origlia11, origlia13, massari14} and Liller 1 \citep[][]{ferraro21, dalessandro22, crociati23, deimer24, fanelli24}. The MGCS targets a large number of still unexplored bulge GC-like systems, and will explore their photometric properties in search of evidence of complex internal evolution that might reveal a similar nature to that of Terzan 5 and Liller 1.\\
{\it Mass function}. The distribution of the individual masses of s stars at their birth defines the so-called initial mass function (IMF). Whether the IMF is universal for all GCs or not has long been debated \citep[][]{bastian10}, but recent investigations seem to indicate that it could depend on the environment in which each cluster was born \citep[][]{henault19, cadelano20}. The IMF is therefore a potential tracer of the origin of a GC. The IMF can be derived from the present-day mass function (PDMF), which differs from the IMF because GCs preferentially lose low-mass stars during their dynamical evolution \citep[e.g.][]{demarchi10}. The simultaneous measurement of the global PDMF and its radial variation makes it possible to reconstruct \citep[][]{webb17, cadelano20} the information on the IMF of the clusters in relative terms (i.e. recognising GCs sharing the same IMF). The MGCS samples the targets' spatial extent typically out to $\sim2$ half-light radii (r$_{h}$), and the simultaneous request for parallel fields in the same bands and with similar depth will enable measurement of the PDMF of the clusters and its radial variation out to $\sim6-7$ r$_{h}$, and thus to recognise GCs that have the same IMF. By coupling this information with the GC orbital properties, the MGCS aims to identify cluster siblings born in the same progenitor galaxy with a completely novel approach. Furthermore, accurate and deep PDMF measurements also enable one to estimate the GC mass and the mass-to-light ratio \citep[][]{baumgardt20}, which are poorly constrained for most of the MGCS targets.

The key objectives listed above are prime examples of a plethora of other science cases of high interest to the community, whose investigation the MGCS will enable. Among the most remarkable, we mention here: $i$) the study of photometric binaries and the dependence of the binary fraction on the different environments of origin \citep[][]{sollima22}; $ii$) the study of blue stragglers as probes of the dynamical processes governing GC evolution \citep[][]{ferraro12}; $iii$) the derivation of new constraints on the structural parameters of the clusters, in particular for the surface density profile of their innermost regions; $iv$) the photometric investigation of chemical peculiarities in GC stellar populations in terms of [Fe/H], He, and C+N+O abundances in the optical CMDs \citep[detected as multiple evolutionary sequences; see e.g.][]{bellini13}; $v$) the improvement of the absolute proper motion determination for each target, which will be achieved by pairing the MGCS observations with first-epoch {\it Gaia} measurements \citep[e.g.][]{massari17, massari18, delpino22}; and $vi$) the provision of invaluable first-epoch measurements for any future proper motion study, possibly in combination with other high-resolution cameras \citep[][]{massari20, haberle21, libralato22, libralato23} that will provide large temporal baselines in the future.

This first paper of the MGCS series is organised as follows. In Sect.~\ref{sec:sample} we present the clusters targeted by the survey and provide key information about their observations. In Sect.~\ref{sec:results} we show the first scientific application, which consists of the age measurement of two surveyed clusters, namely ESO452-11 in the optical and 2MASS-GC01 in the NIR. Lastly, in Sect.~\ref{sec:disc} we discuss the results and draw final conclusions.
%--------------------------------------------------------------------
\section{Selection of the sample and observations}\label{sec:sample}

\begin{figure}[!th]
    \centering
    \includegraphics[width=\columnwidth]{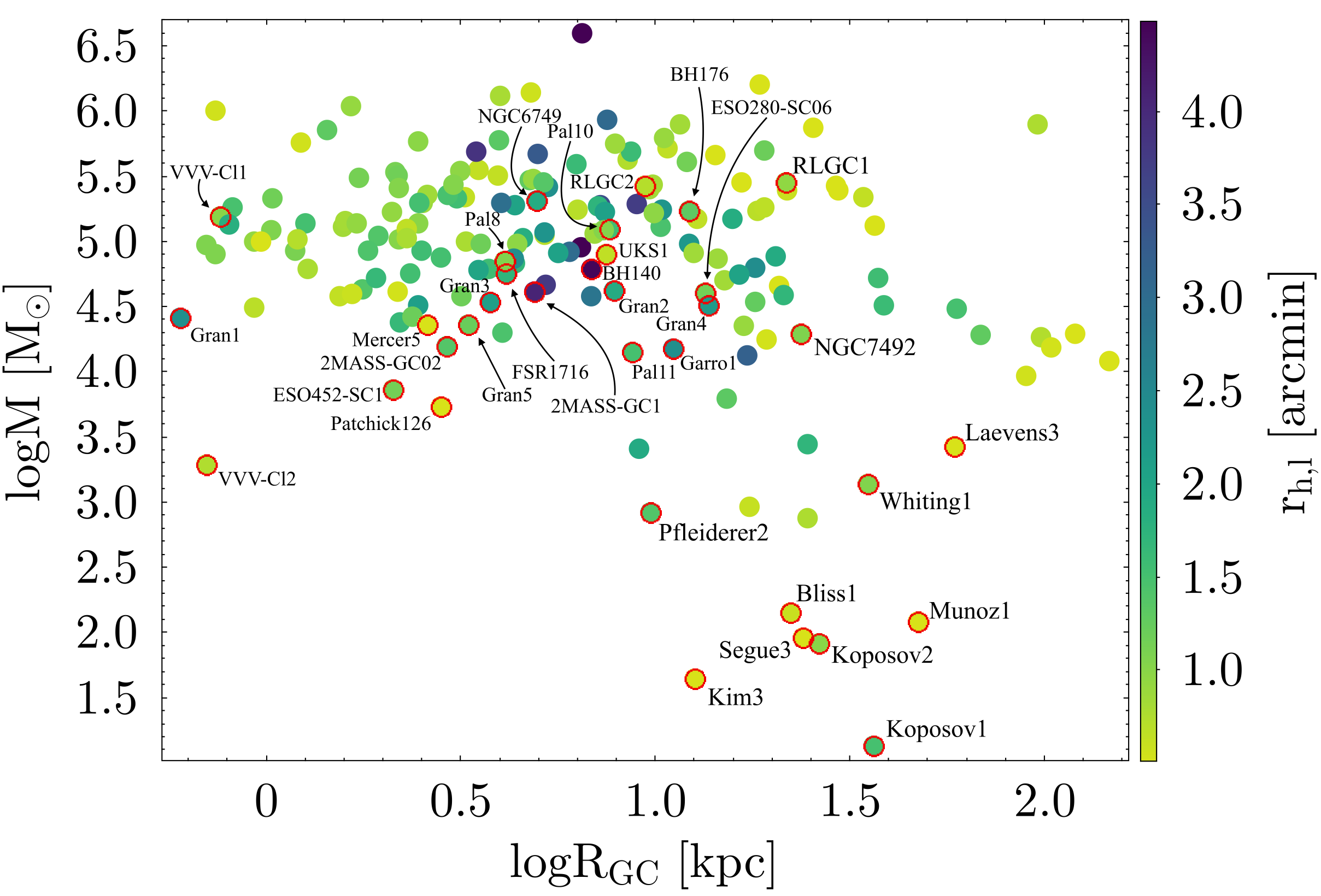}
    \caption{Distribution of the targets of MGCS (labelled and highlighted with red circles) in the Galactocentric distance vs mass plane. Each cluster is colour coded according to its half-light radius.}
    \label{fig:sample}
\end{figure}

The targets of the survey were selected from a number of catalogues, including the compilations by \cite{harris96, vasiliev21, baumgardt21}, and \cite{pace23} and the online database by Holger Baumgardt\footnote{\href{https://people.smp.uq.edu.au/HolgerBaumgardt/globular/orbits.html}{\tt https://people.smp.uq.edu.au/HolgerBaumgardt/globular/\\orbits.html}}. The number of known GC candidates in the MW is steadily increasing \citep[see e.g.][]{garro24, bica24}. To select the targets of the MGCS, we excluded GC candidates that were missing robust kinematic confirmation, and ended up with 34 systems lacking sufficient HST observations to at least build up a CMD.
Most of these 34 systems can be divided into two broad classes (see Fig.~\ref{fig:sample}).
The first is that of low surface-brightness GCs located in the distant halo. The classification as GCs of some of these, such as Koposov-1 and Koposov-2, has been debated \citep[][]{koposov07} and will be addressed as part of the scientific objectives of the survey. 
These 27 targets are located in regions of the sky with sufficiently low reddening that they can be observed with the Wide Field Channel of the ACS (ACS/WFC), following a strategy similar to that defined by \cite{sarajedini07}. We will therefore refer to these targets as the ACS sample. We favoured the ACS camera over the Wide Field Camera 3 (WFC3) because its field of view is $\sim40\%$ larger, and thus it is better suited to investigating the radial behaviour of GC properties.
The chosen F606W and F814W bands were selected because they are perfectly suited for the primary science cases of the survey, as $i$) they minimise the effect that chemical peculiarities in light element
abundances \citep[the multiple populations phenomenon;][]{gratton04, milone22} have on making a GC's CMD more complex, thus increasing the precision of the GC's estimated age; $ii$) they are sensitive to internal differences in age and [Fe/H], and are thus efficient tools for finding evidence of complex internal GC evolution \citep[e.g.][]{bedin04}; and $iii$) they are efficient in sampling very low mass, faint GC stars, in order to explore the mass function to its widest extent \citep{paust10}. The details of the observations of the ACS sample are summarised in Tab.~\ref{tab_acs}.

The second class consists of seven GCs hidden in the highly crowded and extinct regions of the Galactic bulge, where reddening exceeds $E(B-V)>2$. These clusters occupy regions of the parameter space (in terms of mass, density, and evolution environment) that are rather extreme and poorly sampled by existing observations, and where there is room for significant scientific advancement.
These GCs have been imaged with the IR channel of the WFC3, in the F125W and F160W bands. F125W is preferred over the wider F110W because it makes the observations more efficient due to the extreme extinction. These
IR bands can cause the faint main sequence of a GC CMD to split, due to the aforementioned phenomenon of multiple populations \citep{milone17}, but this effect takes place well below the MSTO point and thus does not affect either age estimates or mass-function studies (the mass-luminosity relation along the GC main sequence is not altered significantly). The details of the observations of the WFC3/IR sample are summarised in Tab.~\ref{tab_wfc3ir}.

The MGCS data reduction follows the well-established workflow outlined in various papers focused on high-precision astrometry and photometry with {\it HST} images \citep[e.g.][]{2017BelliniwCenI,nardiello2018,libralato22}, and will be detailed in a forthcoming paper (Libralato et al., in prep.). Briefly, our data reduction consists of two steps called first- and second-pass photometry. First-pass photometry measures the position and flux of bright, relatively isolated sources in each image via effective point spread function (ePSF) fit. For the ACS/WFC and WFC3/UVIS analyses, we made use of the \texttt{\_flc} images\footnote{The \texttt{\_flc} images are a product of the official \textit{HST} pipeline. These exposures have been: (i) dark- and bias-subtracted, (ii) flat-fielded, and (iii) corrected for charge-transfer efficiency defects (i.e. the damaging effects of cosmic rays on the detectors, which affects both the position and flux of stars) as described in \citet{2018acs..rept....4A}. The \texttt{\_flc} images are also not resampled, thus retaining information about where exactly a photon landed on the detector, an essential criterion for achieving high-precision astrometry. In contrast, other \textit{HST} products such as the \texttt{\_drc} exposures, which are resampled images obtained by combining multiple \texttt{\_flc} images, preserve flux but not a source position and/or shape.}, whereas for the WFC3/IR case, {\_flt} images (the analogue of the {\_flc} images, but without the charge-transfer efficiency correction) were employed. The ePSF models were obtained starting from the publicly available\footnote{\href{https://www.stsci.edu/~jayander/HST1PASS/LIB/PSFs/}{https://www.stsci.edu/$\sim$jayander/HST1PASS/LIB/PSFs/}} library {\it HST} ePSFs, and were fine tuned for each exposure. Positions were corrected for geometric distortion using the publicly available\footnote{\href{https://www.stsci.edu/~jayander/HST1PASS/LIB/GDCs/STDGDCs/}{https://www.stsci.edu/$\sim$jayander/HST1PASS/LIB/GDCs/STDGDCs/}} {\it HST} corrections described in various papers \citep{2006acs..rept....1A,2009PASP..121.1419B,2011PASP..123..622B,2016wfc..rept...12A}.

These initial sets of positions and fluxes were used to set up a common reference frame system, where they could be properly combined in the subsequent step. Astrometrically, such a common reference frame system is designed by using the \textit{Gaia}\,DR3 catalogue. This fixes the axis orientations ($x$ pointing to the west and $y$ to the north), scale (40 mas yr$^{-1}$ and 130 mas yr$^{-1}$ for the GCs in the optical and IR samples, respectively), and position of the cluster centre in the pixel-based frame. This centre of the cluster was chosen so as to have only positive pixel-based coordinates. Photometrically, instrumental magnitudes are rescaled to the magnitude of the longest exposure in each camera and filter.

Second-pass photometry is performed with the code \texttt{KS2} \citep[Anderson in prep.;][]{2017BelliniwCenI,nardiello2018,libralato22}. \texttt{KS2} is designed to address the needs of, and overcome the criticalities related to, crowded environments such as GCs (by ePSF subtracting all nearby neighbour priors to the ePSF fit of a source) and to properly characterise faint stars (using all exposures at once in the detection phase).

Together with positions \citep[registered onto the ICRS system thanks to the {\it Gaia} DR3 catalogue;][]{gaiadr3} and magnitudes (calibrated onto the VEGAmag system) for all detected sources in the {\it HST} fields, our astrophotometric catalogues will also contain several diagnostics of the quality of the ePSF fit and the photometry, as well as a parameter that can be used to infer how well the shape of a source resembles that of the ePSF (i.e. a star). In the following, the analyses on ESO452-11 and 2MASS-GC01 were performed by considering only well-measured sources selected by means of criteria similar to those described in \citet{libralato22}, i.e. a mix of percentile-based selections based on the quality of the ePSF fit, magnitude rms, and ``stellarity'' index, the number of images used in the flux measurement, and the impact of the neighbour flux and local sky.

\begin{figure*}
    \centering
    \includegraphics[width=\textwidth]{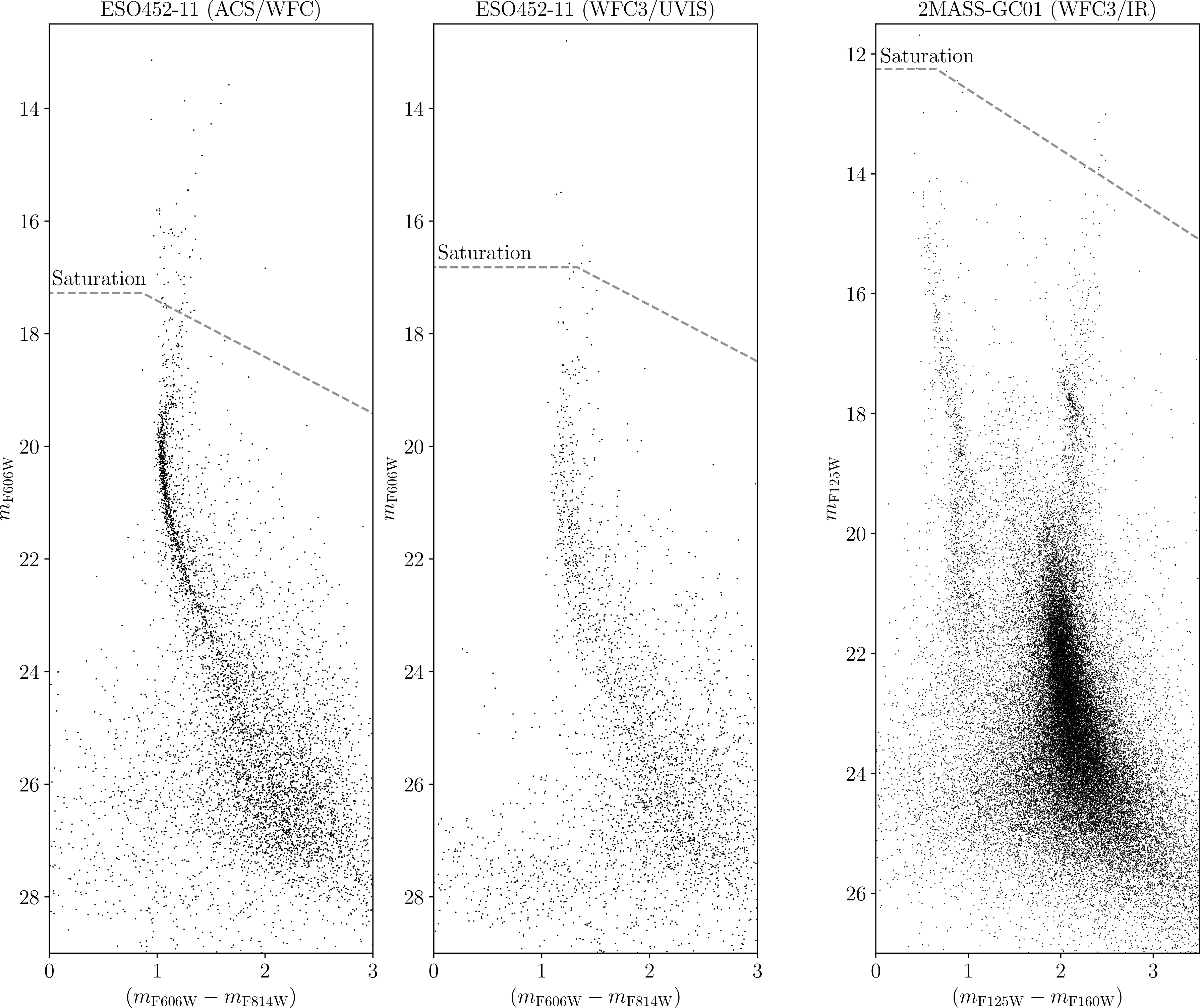}
    \caption{Overview CMDs (not corrected for differential reddening) for ESO452-11 and 2MASS-GC01. Left and middle panels: CMDs of ESO452-11 considering all stars in the central ACS/WFC and parallel WFC3/UVIS fields, respectively. Right panel:  CMD of 2MASS-GC01 based on WFC3/IR data. The dashed grey lines mark the saturation threshold.}
    \label{cmdall}
\end{figure*}

\section{Showcase of scientific applications: Age determination}\label{sec:results}

The scientific applications in this paper focus on two GCs, one for each of the ACS and WFC3/IR samples. The former is ESO452-11 \citep[also known as 1636-283 in][]{harris96}, a cluster discovered by \cite{lauberts81} as part of the European Southern Observatory (ESO) Uppsala survey of the ESO(B) atlas. 
The latter is 2MASS-GC01, an extremely extincted stellar system serendipitously discovered by \cite{hurt2000} from the inspection of the Two Micron All-Sky Survey \citep[2MASS,][]{2mass} data.

\begin{figure*}[!th]
    \centering
    \includegraphics[width=\columnwidth]{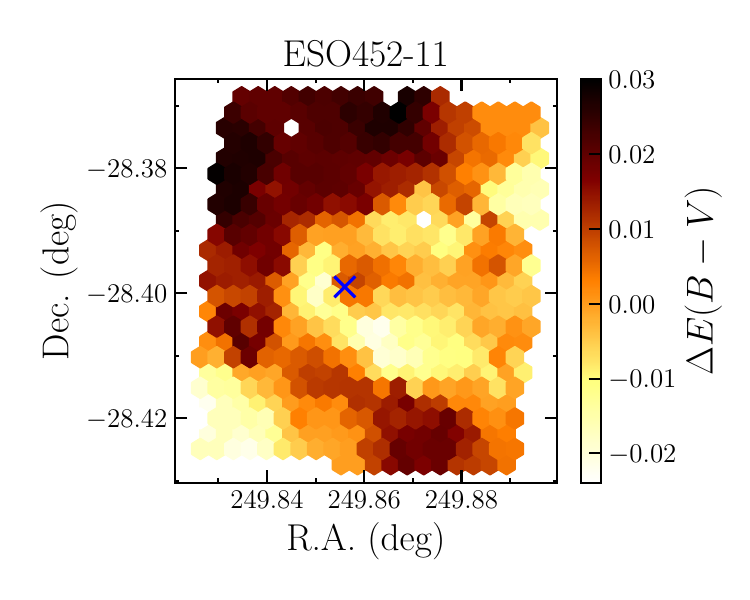}
    \includegraphics[width=\columnwidth]{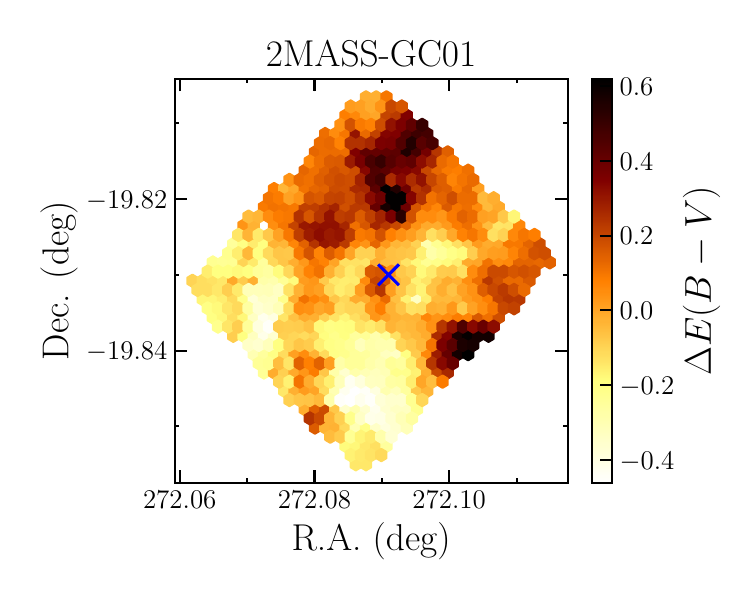}
    \caption{Differential reddening maps for ESO452-11 (left panel) and 2MASS-GC01 (right panel). Please note the very different amplitude of the $\Delta E(B-V)$ scale in the two panels. The centre of the clusters as quoted in \citet[][2010 edition]{harris96} is shown as a blue cross.}   
    \label{redmap}
\end{figure*}

For both systems, we showcase the quality of our MGCS observations by presenting the deepest CMDs (not corrected for differential reddening) currently available in Fig.~\ref{cmdall}. Upon applying quality selection to identify well-measured stars, these CMDs enable the most precise age determination to date for these systems.
We also corrected our photometry for the effects of differential reddening with the method described by \citet{milone12}, before performing the isochrone fitting procedure. The resulting differential reddening maps are shown in Fig.~\ref{redmap} (we refer the reader to Appendix A for a direct comparison between the CMDs before and after correction).

\subsection{Colour-magnitude diagrams}

The left panel of Fig.~\ref{cmdall} shows the CMD of the ACS field approximately centreed on ESO452-11 and covering a radial range between 0.0 and $\simeq 2.0\arcmin$, corresponding to $\simeq 1.7$ r$_h$ \citep[all structural parameters are taken from][]{vasiliev21}. The CMD in the middle panel is from the parallel WFC3 field, sampling a radial range between $4.5\arcmin$ and $7.4\arcmin$. Since the tidal radius of the cluster is $r_t=4.9\arcmin$, this CMD virtually samples only the background population, and the comparison between the two panels clearly reveals the main cluster sequences. From $m_{F606W}\simeq 13.5$ to $m_{F606W}\simeq 19.0$ and around $m_{F606W}-m_{F814W}\simeq 1.2$, we observe a steep, sparsely populated, red giant branch (RGB), with a clumpy red horizontal branch (HB) at $m_{F606W}\simeq 16.3$. The subgiant branch (SGB) is clearly visible, bending from the base of the RGB to the turn-off (TO) point at $m_{F606W}\simeq 20.0$. Below the TO, a thin main sequence (MS) is clearly distinguishable down to $m_{F606W}\simeq 24.5$, where it begins to be overwhelmed by contamination from the Galactic background. The hint of a parallel sequence of cluster binary stars can be perceived in the same magnitude range. The density along the MS is observed to decrease significantly, relative to the TO region, as early as around $m_{F606W}\simeq 22.0$. At this magnitude, completeness should still be very high. \citep{milone12} estimated a completeness of about 90\% for NGC~2298, which was observed with the same instrumental setup and exposure times \citep[e.g.][]{milone12}, and has a slightly higher central stellar density compared to ESO452-11. This might suggest that the low-mass population of cluster stars has been strongly depleted by internal relaxation. The cluster might therefore be in the last phases of its dynamical evolution before disruption \citep{paust10}. Indeed, in Baumgardt's catalogue of GC structural parameters, the reported remaining lifetime (T$_{dis}$) for ESO452-11 is just T$_{dis}$=0.1~Gyr. However, this speculative interpretation requires a more detailed analysis of the cluster dynamics and photometric completeness to be confirmed on a more robust basis.

In the right panel of Fig.~\ref{cmdall}, the CMD of the stars in the WFC3/IR field centred on 2MASS-GC01 is displayed. The field samples the cluster out to $\simeq 1.6\arcmin$ from the cluster centre, which is approximately $1~r_h$ according to \citet{bonatto08} and \citet{harris96}, but $\simeq 0.4~r_h$ according to Baumgardt's catalogue. In any case, our data sample less than half of the cluster light, around the cluster centre. The extended and nearly vertical sequence that is clearly evident in the CMD for $m_{F125W}-m_{F160W}\la 1.3$ is composed of local disc stars ($D\sim 1$~kpc, according to the Gaia DR3 parallaxes of its brightest members). The much redder colour of all the cluster sequences implies a significant increase in the extinction along the line of sight from the location of these stars to that of the cluster ($D\simeq 3$~kpc). The RGB of the cluster is clearly seen around  $m_{F125W}-m_{F160W}\simeq 2.2$, running from $m_{F125W}\simeq 14.0$ to $m_{F125W}\simeq 21.0$, with a red HB at $m_{F125W}\simeq 18.0$. The cluster's upper MS is by far the most prominent feature of the entire CMD between $m_{F125W}\simeq 21.0$ and $m_{F125W}\simeq 26.0$. A prominent blue plume seems to emerge from the old MS in the colour range $1.3\la m_{F125W}-m_{F160W}\la 1.8$. Further investigation is required to establish if this is a pure sequence of blue stragglers or if there is also a (largely sub-dominant) young population as found, for example, in Liller~1 \citep{dalessandro22}. All the cluster sequences are
significantly elongated by the effect of differential reddening (see Fig.\ref{redmap}, right panel).

\subsection{Age estimate}\label{sect:age}

\begin{figure*}
        \centering
        \begin{subfigure}[b]{0.45\textwidth}
            \centering
            \includegraphics[width=\textwidth]{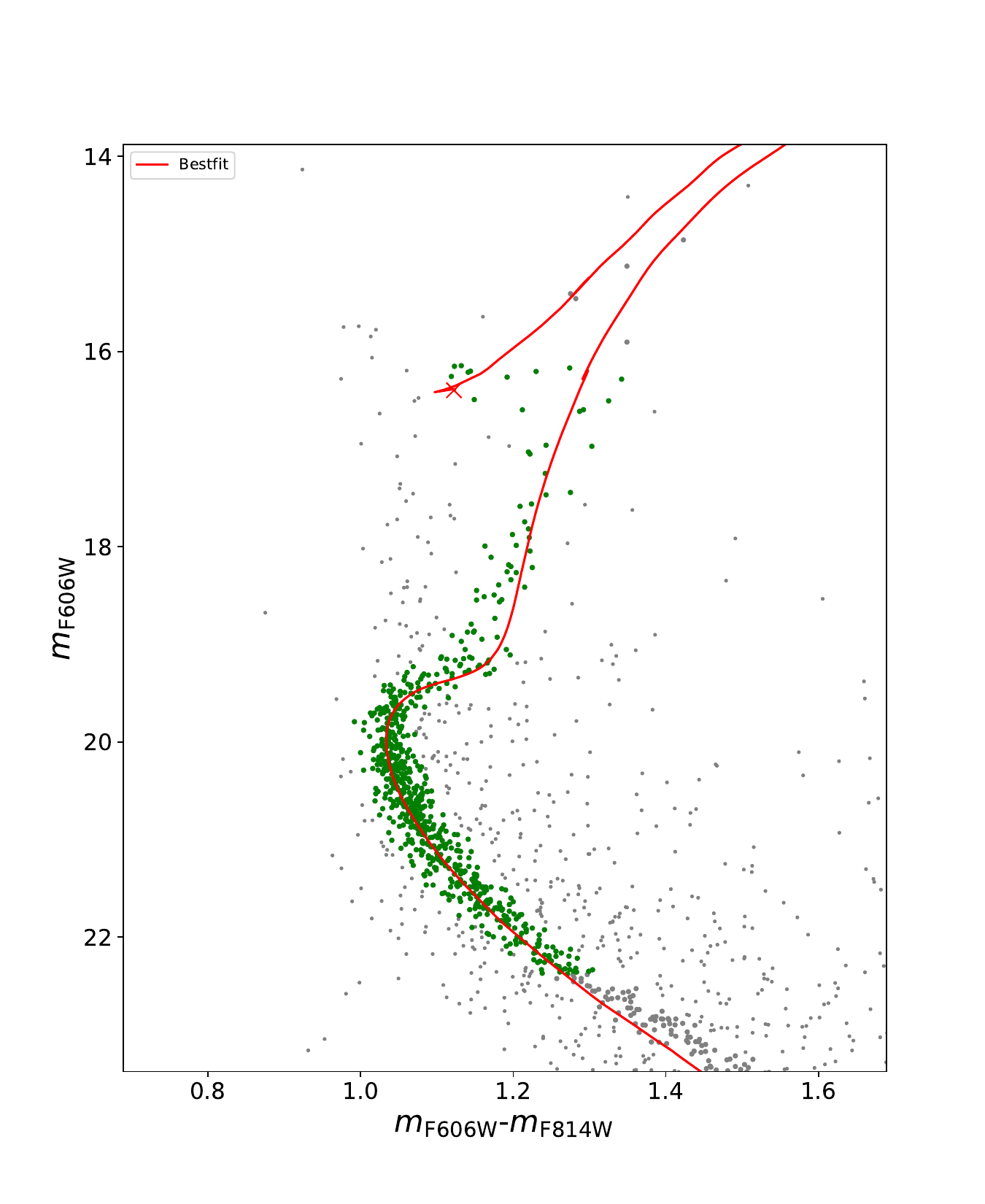}
            \caption[]%
            {{\small }}    
            %\label{fig:mean and std of net14}
        \end{subfigure}
        \hfill
        \begin{subfigure}[b]{0.45\textwidth}  
            \centering 
            \includegraphics[width=\textwidth]{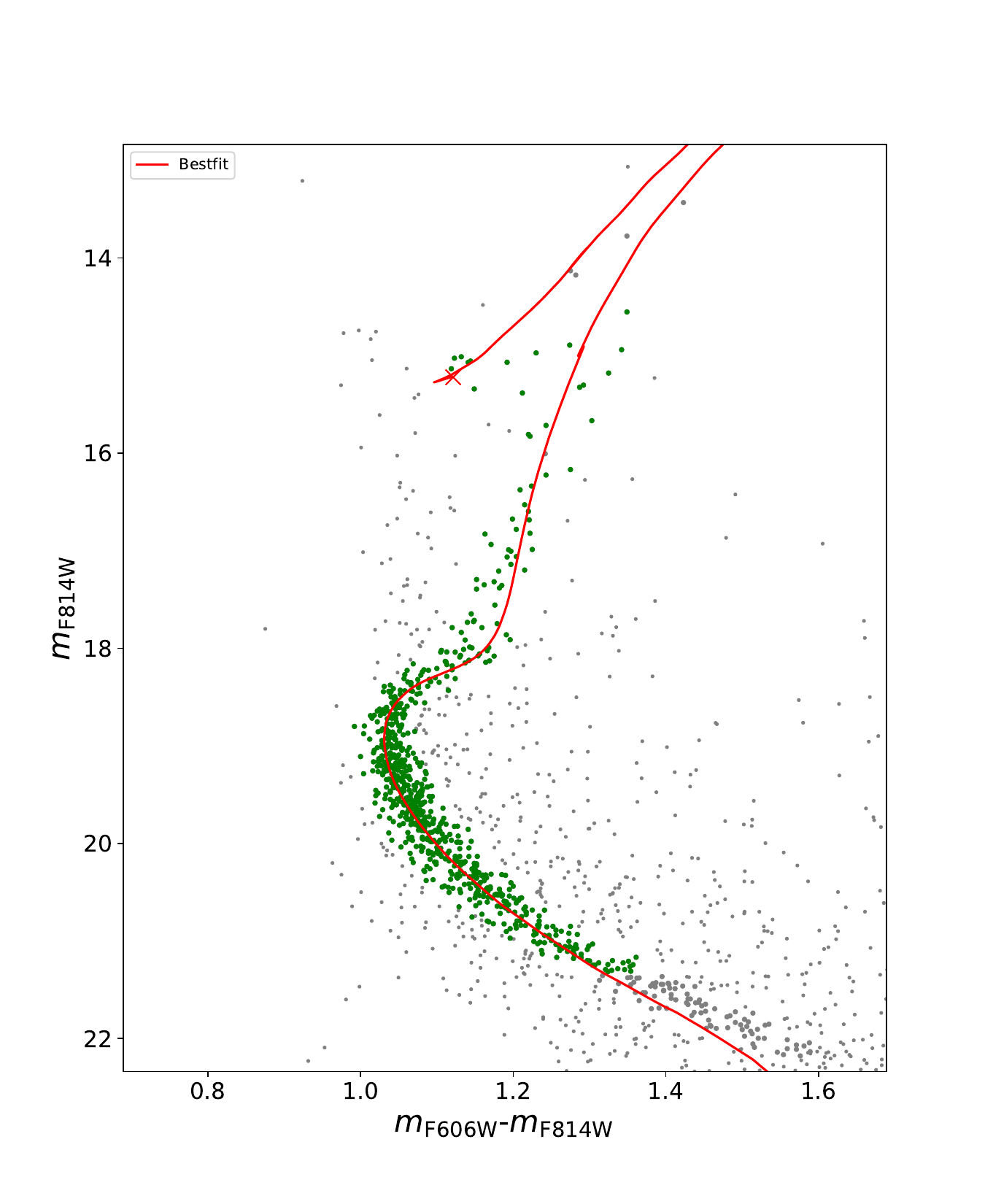}
            \caption[]%
            {{\small }}    
            %\label{fig:mean and std of net24}
        \end{subfigure}
        \vskip\baselineskip
        \begin{subfigure}[b]{0.45\textwidth}   
            \centering 
            \includegraphics[width=\textwidth]{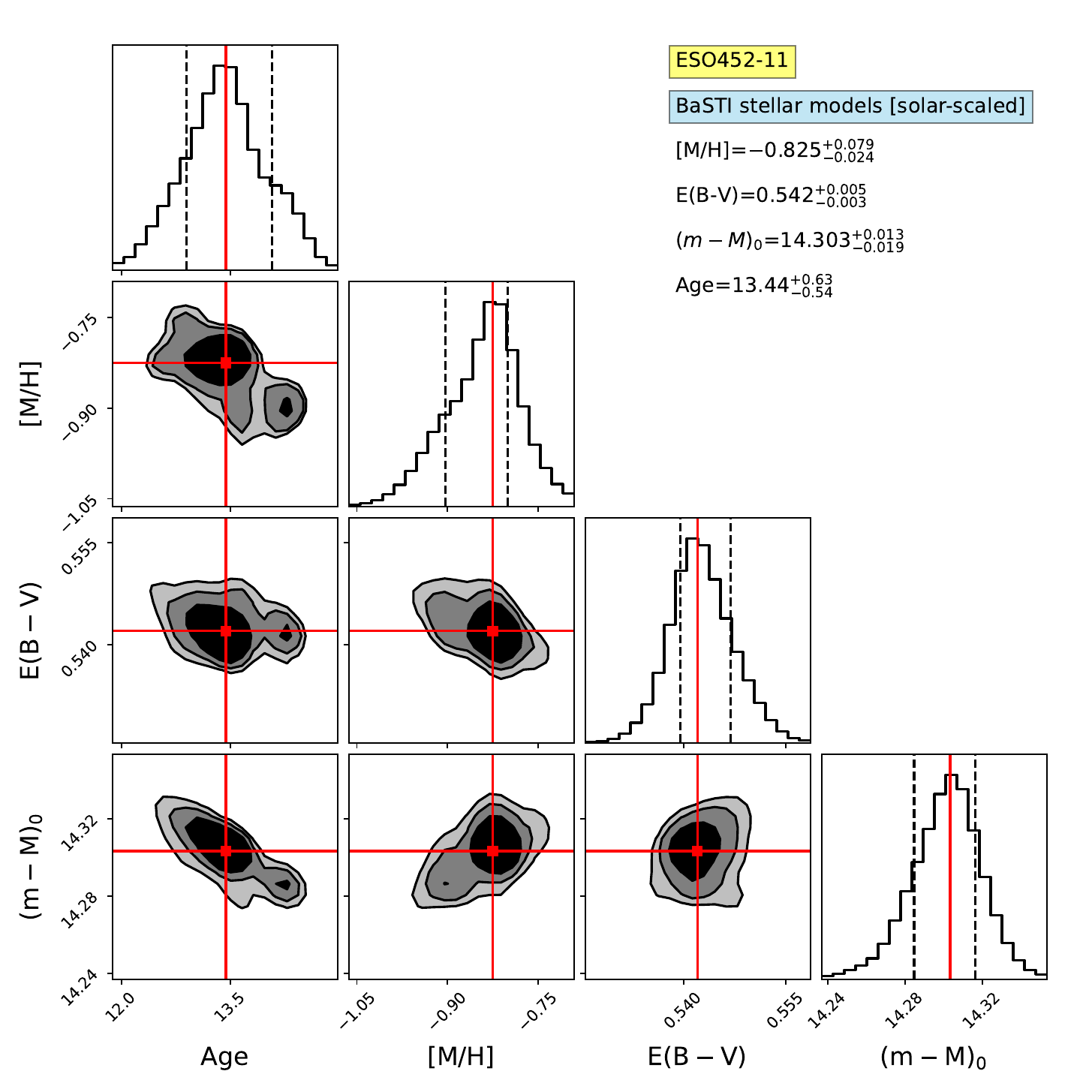}
            \caption[]%
            {{\small }}    
            %\label{fig:mean and std of net34}
        \end{subfigure}
        \hfill
        \begin{subfigure}[b]{0.45\textwidth}   
            \centering 
            \includegraphics[width=\textwidth]{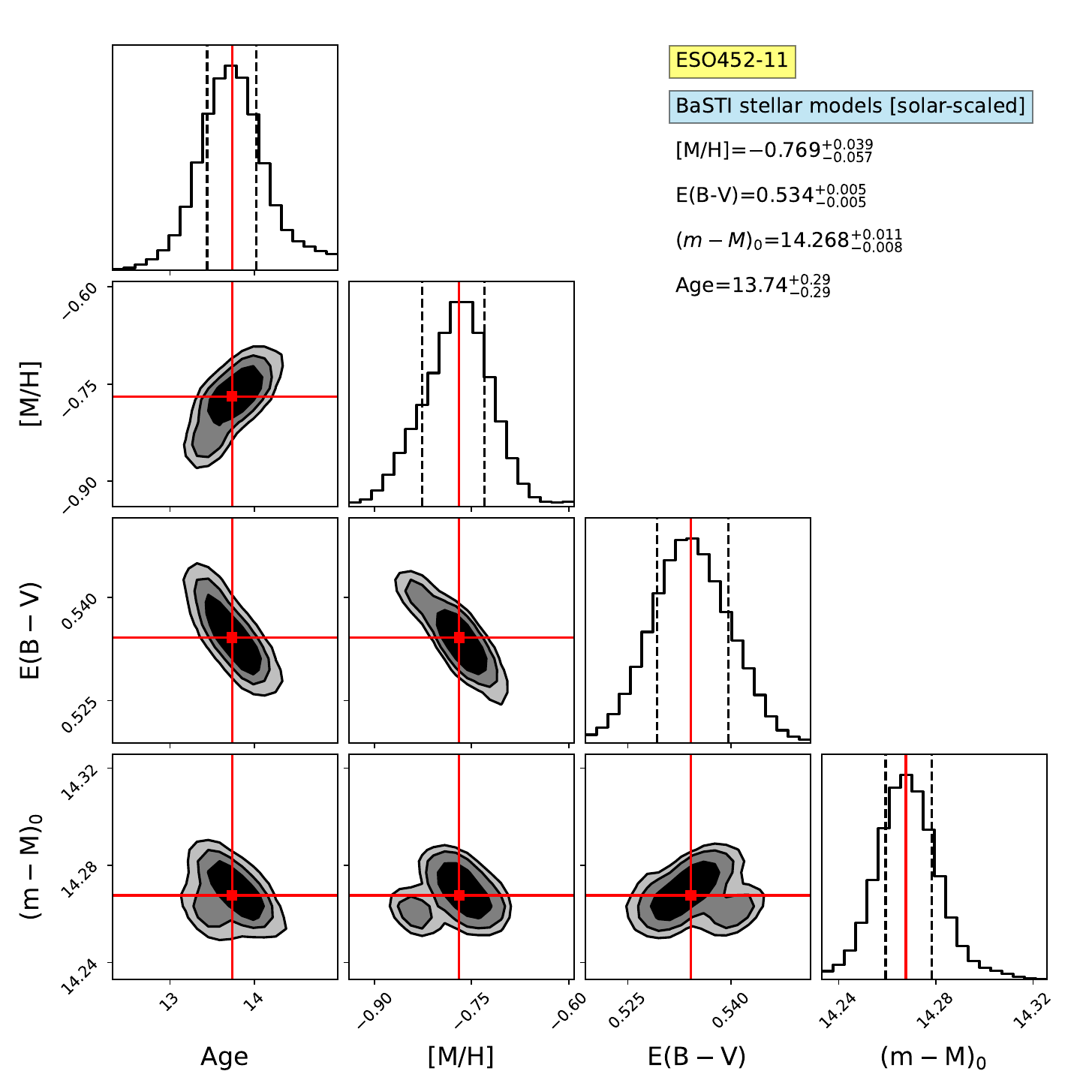}
            \caption[]%
            {{\small }}    
            \label{fig:mean and std of net44}
        \end{subfigure}
        \caption[]
        {\small Results of the isochrone fitting for ESO452-11. Panels a) and b) show the ($m_{\rm F606W}-m_{\rm F814W}$, $m_{\rm F606W}$) and the ($m_{\rm F606W}-m_{\rm F814W}$, $m_{\rm F814W}$) CMDs, respectively. The grey symbols describe the whole sample of stars in the cluster catalogue, while green symbols show the stars effectively used for the isochrone fit, selected as described in the text. The red line corresponds to the best-fit model. Panels c) and d) show the corresponding corner plots of the MCMC sampling.} 
        \label{age_eso452}
    \end{figure*}

\begin{figure*}
        \centering
        \begin{subfigure}[b]{0.45\textwidth}
            \centering
            \includegraphics[width=\textwidth]{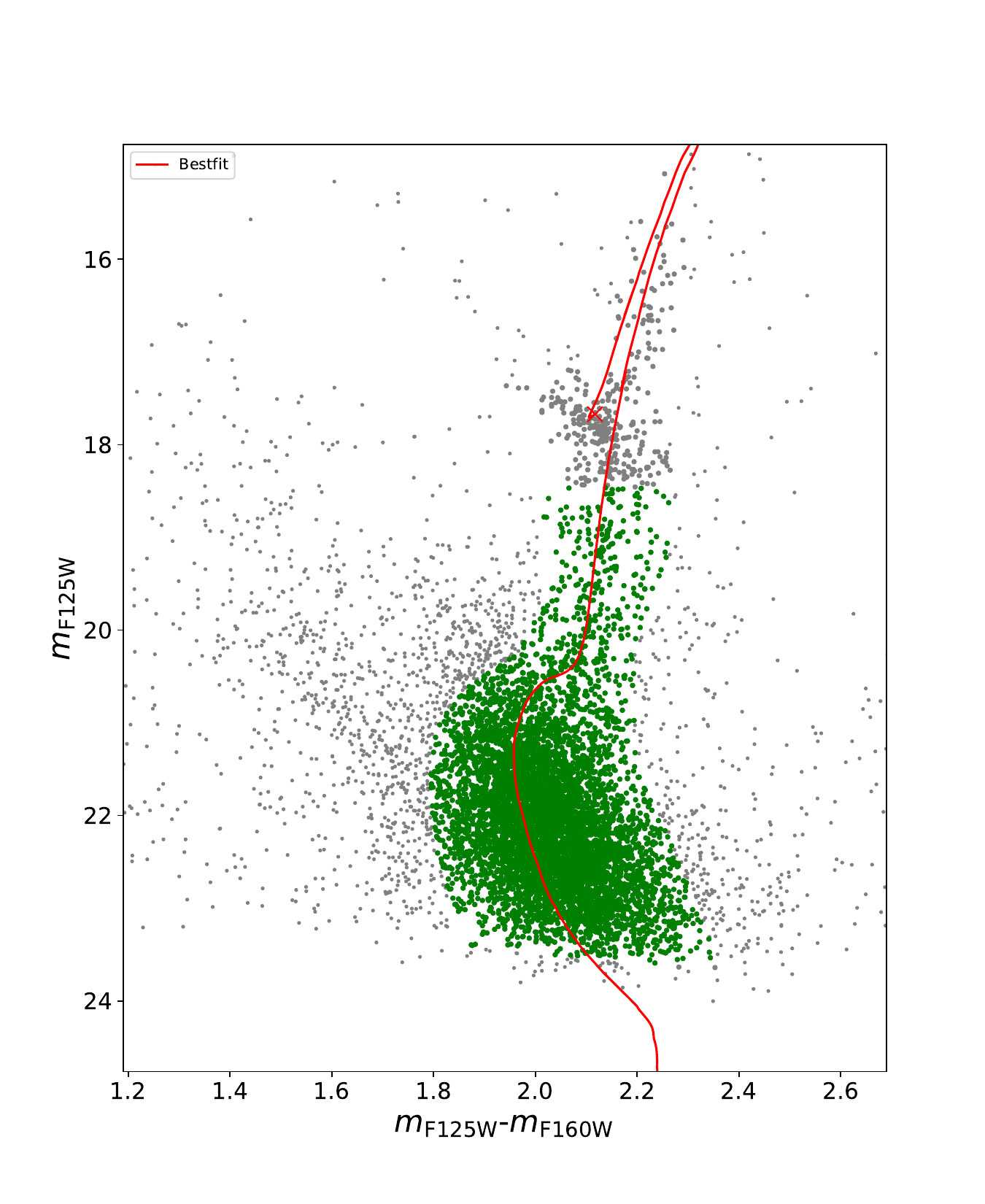}
            \caption[]%
            {{\small }}    
            %\label{fig:mean and std of net14}
        \end{subfigure}
        \hfill
        \begin{subfigure}[b]{0.45\textwidth}  
            \centering 
            \includegraphics[width=\textwidth]{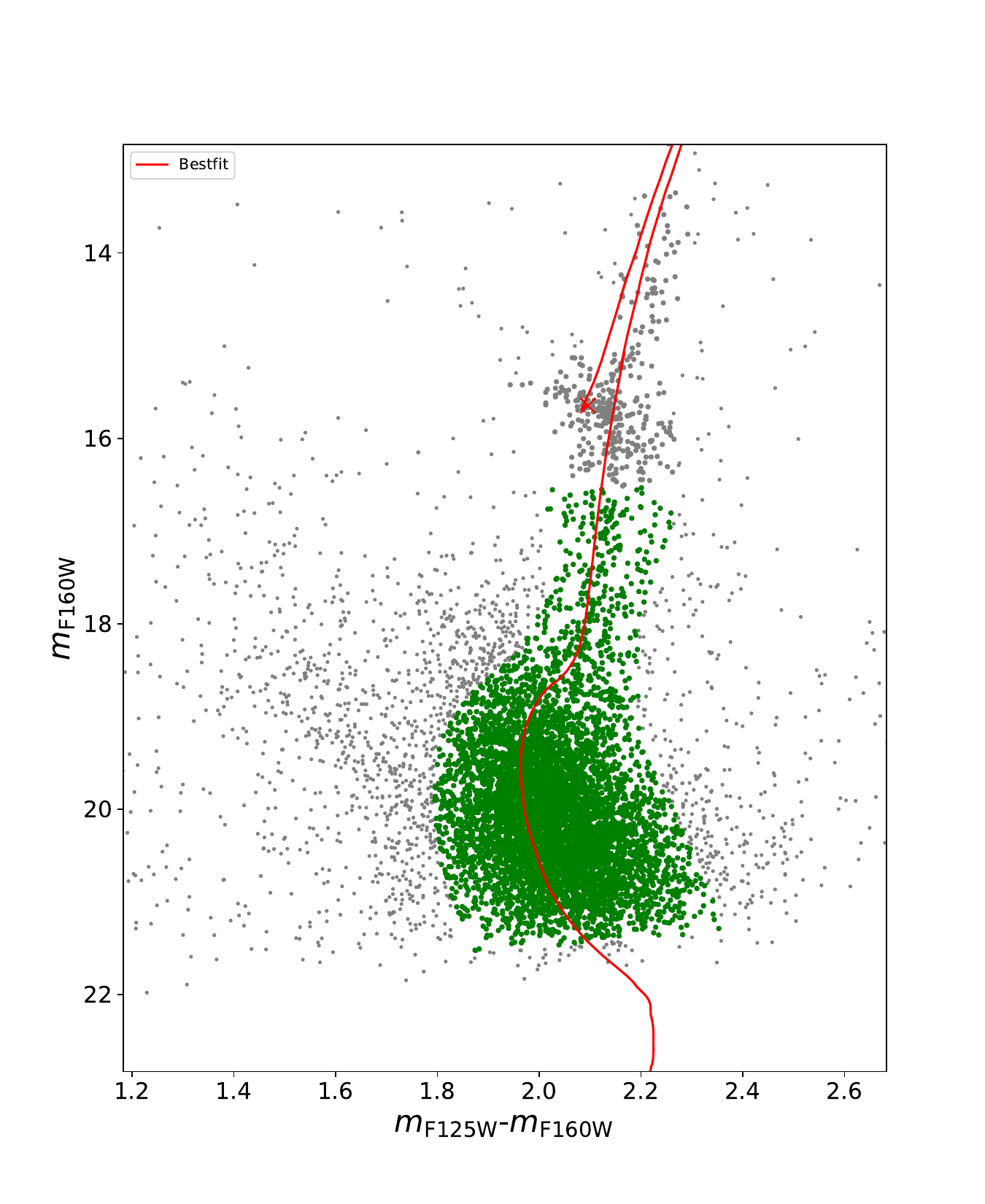}
            \caption[]%
            {{\small }}    
            %\label{fig:mean and std of net24}
        \end{subfigure}
        \vskip\baselineskip
        \begin{subfigure}[b]{0.45\textwidth}   
            \centering 
            \includegraphics[width=\textwidth]{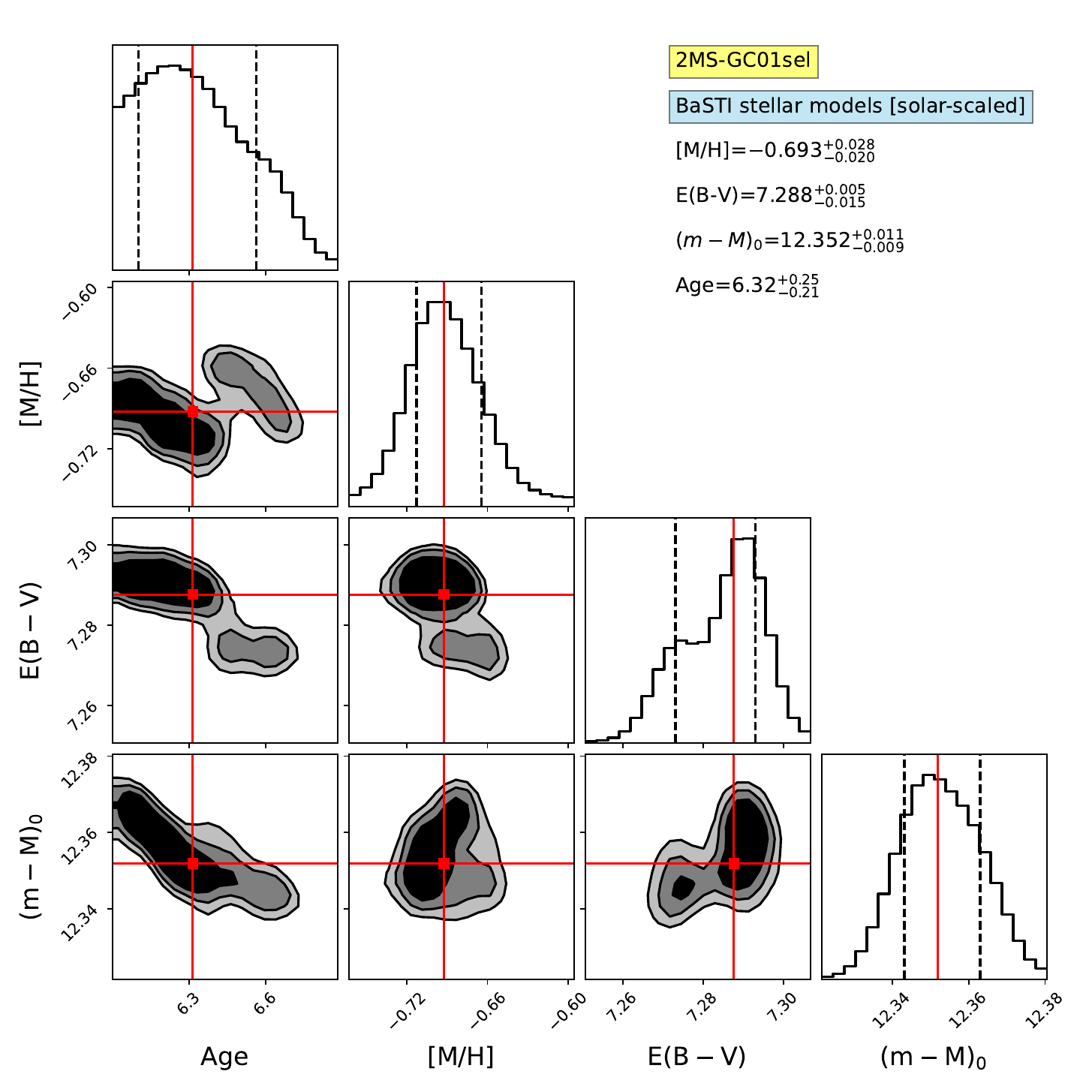}
            \caption[]%
            {{\small }}    
            %\label{fig:mean and std of net34}
        \end{subfigure}
        \hfill
        \begin{subfigure}[b]{0.45\textwidth}   
            \centering 
            \includegraphics[width=\textwidth]{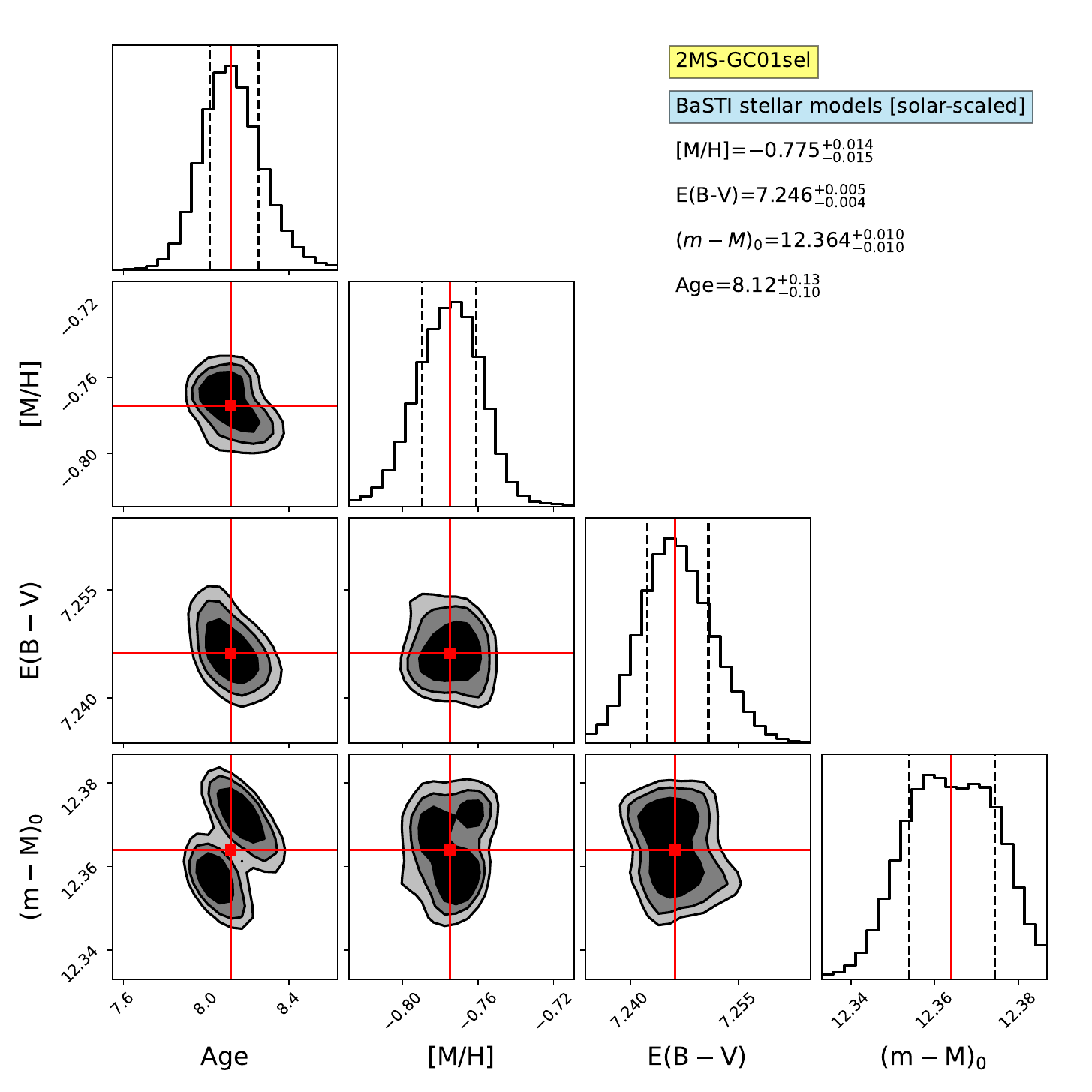}
            \caption[]%
            {{\small }}    
            \label{fig:mean and std of net44}
        \end{subfigure}
        \caption[]
        {\small Results of the isochrone fitting for 2MASS-GC01. } 
        \label{age_2mass}
    \end{figure*}

The MGCS takes advantage of the methods and procedures developed within the CARMA project \citep{massari23}, which aims at determining the age of the entire sample of Galactic GCs in the most homogeneous way. In this sense, the MGCS perfectly fits the aims of CARMA, as the {\it HST} photometry is obtained in the same reference photometric system as that selected by CARMA, which is the one defined by the ACS filters. To briefly summarise the CARMA methods: our age determination followed an isochrone fitting approach that provided the best-fit values for age, metallicity, distance, and colour excess within a Bayesian statistical framework, thus associating robust uncertainties with all of the output parameters.
In particular, the fitting procedure started from defining the mean ridge line of the differential-reddening corrected CMD, and used this to select the sample of stars to fit. The selection included all stars within a certain colour range from the ridge line (see the green symbols in Fig.~\ref{age_eso452} and Fig.~\ref{age_2mass}), and was applied to exclude obvious stellar contaminants of the field, and photometric binaries, whenever the photometric quality allowed it. A fine grid of solar-scaled theoretical models from the BaSTI stellar evolution library \citep{hidalgo18, pietrinferni21} was then fitted\footnote{The goodness of the fit was evaluated through a likelihood composed of two terms, one related to the distance of each star to the isochrone, and the other to the consistency of the inferred parameters with the initial priors (see Section 2.2 of \citealt{massari23} for details).} to the data via Markov chain Monte Carlo (MCMC) sampling, starting from some loose priors on metallicity, distance, and $E(B-V)$. The choice of solar-scaled models was made to avoid making any assumptions on the $\alpha$-element abundance of GCs. The term on the $\alpha$-element mixture is then effectively absorbed by working in global metallicity [M/H]. This is justified by the fact that the MGCS photometry is in the optical and NIR bands, for which the equivalency between solar-scaled and $\alpha$-enhanced models at the same global metallicity has been demonstrated by \cite{salaris93} and \cite{cassisi04} through the relation $[M/H] =[Fe/H]+\log(0.694\times10^{[\alpha/Fe]}+0.301)$. During the fitting procedure, the magnitudes in each passband of the theoretical models were corrected for temperature-dependent extinction, following the prescriptions of \cite{girardi08}. Finally, the median values of the resulting posterior distribution functions were used as the best-fit solution, and the 16th and 84th percentiles were taken as the asymmetric uncertainties. All the details of the method can be found in \cite{massari23}. 

\begin{table}[htp]
    \centering
    \begin{tabular}{|c|c|c|}
    \hline
    {Prior parameter} & {ESO452-11}  &  {2MASS-GC01}  \\ 
    \hline
    [M/H] & -0.9 & -0.7 \\ 
    $\sigma_{{\rm M/H}}$ & 0.1 & 0.1 \\ 
    E(B-V) & 0.55 & 6.8 \\ 
    $\sigma_{{\rm E(B-V)}}$ & 0.05 & 0.20 \\ 
    DM$_0$ & 14.40 & 12.50 \\ 
    $\sigma_{{\rm DM_0}}$ & 0.10 & 0.10 \\ 
    \hline
    \end{tabular}
    \caption{Priors adopted for the isochrone fit.}
    \label{tab:my_label}
\end{table}

Here we show the age determination results for the pair of GCs analysed in this work, starting with ESO452-11. From the photometric point of view, shallow, ground-based, optical and NIR observations of this cluster have been suggested \citep{minniti95, bica99, cornish06, bonatto08}. ESO452-11 is 10--16 Gyr old, with a reddening of $E(B-V)=0.5$--0.8 mag, a distance of 5.8--7.5 kpc, and a photometric metallicity of  $-1.3$$<$[Fe/H]$<$$-0.4$. Recent spectroscopic investigations \citep{koch17, simpson17} determine a more precise mean metallicity of [Fe/H]$=-0.88\pm0.03$, with a possible spread in light elements \citep{simpson17} and a slight underabundance of the $\alpha$-elements compared to the surrounding bulge stars \citep{koch17}. This latter chemical feature might suggest a different origin for ESO452-11 relative to in situ stars. Determining a precise age and comparing that with the age of other in situ GCs of similar metallicity will provide independent insight into the origin of ESO452-11.

Figure \ref{age_eso452} shows the results of the fit in the ($m_{\rm F606W}-m_{\rm F814W}$, $m_{\rm F606W}$) and ($m_{\rm F606W}-m_{\rm F814W}$, $m_{\rm F814W}$) planes, independently. The CMDs and the best-fit isochrones are shown in the upper row, while the marginalised two- and one-dimensional distributions of the model parameters are shown in the bottom row. 
The two solutions are mutually consistent in all the parameters of the fit. For the final results, we used the average values of the two solutions, while the overall uncertainties were computed so as to encompass the upper and lower limits of both runs combined. 

The mean colour excess is $E(B-V)=0.54\pm0.01$, consistent with the previous estimates. The mean distance modulus is $(m-M)_0=14.29\pm0.03$, corresponding to a heliocentric distance of $7.2\pm0.1$ kpc, consistent within 1$\sigma$ with \cite{baumgardt21}, who find $7.39\pm0.20$ kpc, and in agreement with the range from 6.6 kpc to 7.5 kpc quoted by \cite[][]{cornish06}. Note, however, that adopting the integrated apparent $V$ magnitude $V_T=11.77\pm 0.09$ from \citet{baumgardt20} and our estimates of the distance modulus and reddening, the absolute magnitude of the cluster brightens from $M_V=-3.82$, as reported by \citet{baumgardt20}, to $M_V=-4.19\pm 0.10$. The mean global metallicity is ${\rm [M/H]}=-0.80^{+0.08}_{-0.11}$. Before comparing this to spectroscopic values, it is essential to convert [M/H] to [Fe/H], correcting by the contribution of $\alpha$-elements \citep{salaris02}. Using Eq.~3 from \cite{salaris93}, adjusted by the different reference solar mixture of the BaSTI models \citep[see e.g.][]{massari23} and adopting the mean ${\rm [(Si+Ca+Ti)/Fe]}=0.20$ found for this cluster by \cite{koch17}, our solution translates to ${\rm [Fe/H]}=-0.94$. This is significantly different from the value of [Fe/H]$=-1.50$ quoted by \cite{harris96}, but consistent within 1$\sigma$ with the two most recent spectroscopic investigations available in the literature \citep[][]{koch17, simpson17}.

The mean age of ESO452-11 is ${\rm t}=13.59^{+0.48}_{-0.69}$ Gyr. Given the systematic errors in the absolute age values introduced by different assumptions on the input physics of theoretical models \citep[e.g.][]{chaboyer95}, we are mainly interested in the relative age of this GC. This can be compared to age values obtained in a strictly homogeneous way, such as the ones provided by CARMA. In this sense, the cluster is older than the in situ GCs at ${\rm [M/H]}\simeq-0.5$ studied by \cite{massari23} and it is as old as the in situ GC NGC~288 at metallicity of ${\rm [M/H]}=-1.11$ studied by \cite{aguado-agelet25}, for which the authors found ${\rm t}=13.75^{+0.28}_{-0.22}$ Gyr. Therefore, these results point towards an in situ origin for ESO452-11 and support its classification as a bulge GC based on its orbital properties as proposed by \citep[][eDR3 edition\footnote{\href{https://www.oas.inaf.it/en/research/m2-en/carma-en/}{\tt https://www.oas.inaf.it/en/research/m2-en/carma-en/}}]{massari19} and \citet{callingham22}.

The second cluster presented here, 2MASS-GC01, is a rather faint system \citep[$M_V=-5.8$;][]{bonatto08} hidden behind thick dust clouds in the disc. For this reason, the cluster has been very poorly investigated prior to this work. The only information available comes from \cite{ivanov02}, who determined a photometric metallicity of ${\rm [Fe/H]}=-1.19\pm0.38$ and an extinction $E(B-V)\sim6.8\pm0.71$.

Our photometry is by far the deepest ever obtained for 2MASS-GC01, sampling $\sim2$ mag below the cluster MSTO, whereas previous photometry could only detect the cluster RGB. This provides a unique opportunity to reliably estimate the age of the cluster.
As evident from the shape of the red clump in the CMD of Fig.~\ref{age_2mass}, despite the differential reddening correction, the photometry still suffers from severe residual differential reddening. The age estimate is therefore more uncertain than in the case of ESO452-11, but our code still converges to reasonable solutions for the ($m_{\rm F125W}-m_{\rm F160W}$, $m_{\rm F125W}$) and ($m_{\rm F125W}-m_{\rm F160W}$, $m_{\rm F160W}$) CMDs, with the isochrone fit well able to reproduce the most evident features. 

In terms of average values, our combined solution points to a metallicity of ${\rm [M/H]}=-0.73^{+0.06}_{-0.06}$.
Since no spectroscopic measurements of the $\alpha$-element abundance exist for this cluster, but its global metallicity is similar to the one found for ESO452-11, and both clusters are likely born in situ, we assumed the same [$\alpha$/Fe] ratio\footnote{An uncertainty of $\pm0.15$ in [$\alpha$/Fe] translates to an additional uncertainty of $\pm0.11$ in [Fe/H].}, thus obtaining ${\rm [Fe/H]}=-0.85^{+0.07}_{-0.06}$. This value is consistent within 1$\sigma$ with the photometric metallicity derived by \cite{ivanov02}. The average colour excess and distance modulus are $E(B-V)=7.27^{+0.02}_{-0.03}$ and $(m-M)_0=12.36^{+0.03}_{-0.02}$, respectively. These values are also in agreement within the (rather large) uncertainty of previous literature estimates and make 2MASS-GC01 the most extincted cluster known in the Milky Way. It is located at a distance of $D=2.96\pm0.04$ kpc from us, consistent within 1$\sigma$ with $D=3.37\pm 0.62$ kpc from \cite{baumgardt21}. However, we find a mean reddening significantly larger than that found in the literature. Assuming $V_T=27.7$ from Baumgardt's catalogue, this leads to a much brighter absolute magnitude $M_V=-7.20\pm 0.50$ (by arbitrarily adopting an uncertainty of 0.5 mag on $V_T$, since it is not reported in the catalogue). 

In terms of relative age, when compared to the other Milky Way GCs for which age has been estimated in the same homogeneous way,  2MASS-GC01 is very young, with an average of ${\rm t}=7.22^{+0.93}_{-1.11}$ Gyr. For reference, ESO452-11, for which we find a similar metallicity, is $\sim6.5$ Gyr older. The in situ GCs analysed by \cite{massari23} are also older by more than 5 Gyr. As shown by \cite{marin-franch09}, a younger age at this metallicity could be indicative of an accreted origin. Within the CARMA collaboration, \cite{aguado-agelet25} estimated the age of GCs associated with the Gaia-Sausage-Enceladus merger event \citep[][]{helmi2018, belokurov2018}, finding that at ${\rm [Fe/H]}\simeq-1$ these GCs are younger than their in situ counterpart by 1.5--2 Gyr, but not nearly as young as 2MASS-GC01.
The location of 2MASS-GC01 on the age-metallicity plane is instead more consistent with that described by Milky Way disk stars during a late, quiescent phase of star formation, as \cite{xiang22} find. Incidentally, these authors find such a late phase to start at an age of about 8 Gyr, which is similar to the age we estimate for 2MASS-GC01 (but we remark once again that these absolute values should be compared with great caution).

\subsection{The nature of 2MASS-GC01}

\begin{figure}[!th]
    \centering
    \includegraphics[width=\columnwidth]{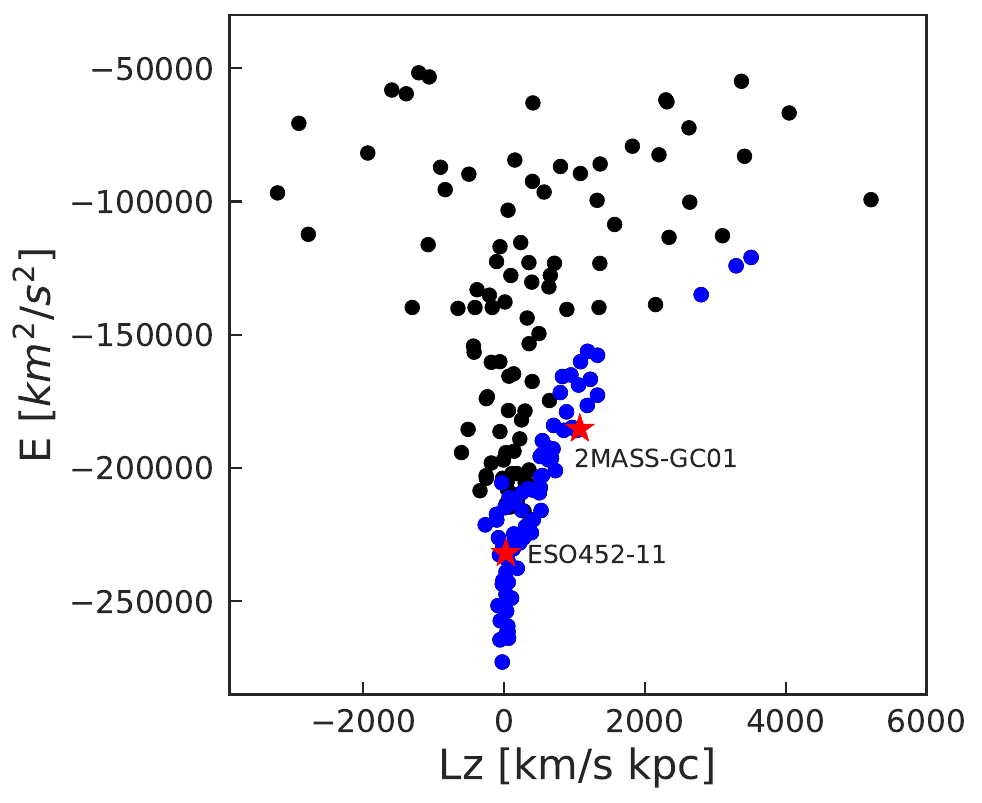}
    \caption{Location in the energy-vertical angular momentum space of 2MASS-GC01 and ESO452-11 (red symbols), compared with the distribution of Mlky Way in situ (blue symbols) and accreted (black symbols) GCs, according to \cite{massari19}, eDR3 edition.}\label{iom}
\end{figure}

The findings presented here could be interpreted as evidence that 2MASS-GC01 is an open cluster (OC) rather than a GC. On the one hand, this conclusion is fully consistent with the dynamical properties of the cluster. In fact, by integrating its orbit into the MW gravitational potential of \citet{mcmillan17}, with the prescriptions listed by \citet{massari19} and \citet{ceccarelli24b}, and by assuming the distance determined here and the remaining five parameters of the phase space as in \citet{baumgardt21}, we find that 2MASS-GC01 moves on a very circular, disc-like orbit. Its circularity is 0.98, with a maximum height from the Galactic plane of only z$_{max}=0.11$ kpc and an eccentricity of 0.14 (resulting from an orbital pericentre of 4.19 kpc and an apocentre of 5.56 kpc). These orbital parameters and the cluster location in the integrals of motion space described by energy and vertical angular momentum (shown in Fig.~\ref{iom}, together with those of ESO452-11) strongly suggest an in situ origin in the Galactic thin disc \citep[in agreement with][eDR3 edition]{massari19}.

On the other hand, its structural parameters are not consistent with those of other typical Galactic OCs. For example, its core radius is $r_{\rm c}=0.18$ pc \citep[][]{baumgardt18}, which is below the lower limit of the OC distribution studied by \citet{tarricq22}. Similarly, its half-light radius is the smallest among the old OCs that \citet{alvarez24} investigate. Its estimated mass of $4.1\times 10^4$ M$_{\odot}$ \citep[][]{baumgardt18} is at the very high mass end of the mass distribution of Galactic OCs \citep[][]{cordoni23, just23, hunt_reffert24}. It is even more extreme if we redetermine it by using the same mass-to-light ratio as in \citet{baumgardt18} and the absolute magnitude derived here. In fact, this would increase the mass estimate to $1.1\times 10^5$ M$_{\odot}$. Ultimately, it should be noted that the metallicity of 2MASS-GC01 is $>0.5$~dex more metal poor than open clusters at similar Galactocentric distances, also considering only OCs older than 5~Gyr \citep{magrini23,palla24}. 

In conclusion, 2MASS-GC01 either looks like a peculiarly young GC orbiting in the thin disc, or a rather extreme OC in terms of mass and compactness. In the absence of a spectroscopic follow-up that can unveil the details of its chemical composition, we limit our conclusions by highlighting the possibility that the classification of 2MASS-GC01 \citep[already identified as a candidate GC by][]{hurt2000} might have to be revisited.

\section{Summary and conclusions}\label{sec:disc}

In this first paper of the series we have presented the {\it Hubble} MGCS, an imaging survey dedicated to the 34 kinematically confirmed MW GCs that lacked prior {\it HST} observations to build a CMD.
After describing the ACS sample, consisting of 27 GCs observed in the F606W and F814W filters of the ACS/WFC, and the WFC3/IR sample, consisting of 7 GCs located in highly extincted regions of the bulge and observed through the F125W and the F160W filters of the WFC3/IR, we demonstrate the quality of the MGCS data with two targets. In particular, we took advantage of the deep and accurate photometry obtained from the observations of ESO452-11 and 2MASS-GC01 to provide the first precise age determinations of the two clusters.

By adopting the tools and the methods developed within the CARMA project \citep{massari23}, we find ESO452-11 to be a $13.59^{+0.48}_{-0.69}$-Gyr-old GC at an intermediate metallicity of ${\rm [M/H]}=-0.80$. These values make ESO452-11 as old as the in situ GC NGC~288 at ${\rm [M/H]}\simeq-1.11$ \citep[][]{ceccarelli25}, and about 1 Gyr older than the in situ GCs at ${\rm [M/H]}\simeq-0.5$ that \citet{massari23} investigate. These results thus confirm the in situ origin of this system, as previous works based on its orbital properties already suggest \citep[e.g.][]{massari19, callingham22}.

In contrast, the isochrone fitting procedure applied to the CMD of 2MASS-GC01 reveals a remarkably young age of ${\rm t}=7.22^{+0.93}_{-1.11}$ Gyr at a metallicity ${\rm [M/H]}=-0.73$. This age is significantly younger than the youngest among the accreted GCs belonging to the Gaia-Sausage-Enceladus merger event \citep[][]{aguado-agelet25}, and is more consistent with the start of the late phase of quiescent star formation in the MW disc, as \citet{xiang22} find. Furthermore, we computed the cluster orbit using the distance estimate derived from our isochrone fitting, and find an almost circular orbit with a very limited excursion above the Galactic plane. For these reasons, our findings suggest that 2MASS-GC01 might be a massive OC rather than a GC. However, the metallicity derived here highlights the cluster as an extreme outlier of the Galactic metallicity gradient as traced by OCs. Moreover, its structural parameters are significantly different from the typical values observed in other OCs. Hence, we prefer to wait for an adequate and detailed spectroscopic abundance analysis before drawing firm conclusions regarding the nature of this system.

The simple scientific applications of the MGCS dataset shown in this paper demonstrate the power these {\it HST} observations have to push our knowledge on GCs beyond its current limits. The results of the isochrone fitting of MGCS GCs data will be kept up to date at the CARMA public repository\footnote{\href{https://www.oas.inaf.it/en/research/m2-en/carma-en/}{\tt https://www.oas.inaf.it/en/research/m2-en/carma-en/}}.
Calibrated and geometric distortion corrected astrophotometric catalogues including quality parameters for an appropriate sample selection, as well as differential reddening maps and catalogues of artificial stars for each target of the survey will be released to the public, thus completing the legacy of {\it HST} in this paramount field of astronomical research.

\begin{acknowledgements}
We thank the anonymous referee for their thoughtful suggestions on how to improve the quality of our paper.
DM, SC, EP and AM acknowledge financial support from PRIN-MIUR-22: CHRONOS: adjusting the clock(s) to unveil the CHRONO-chemo-dynamical Structure of the Galaxy” (PI: S. Cassisi). SC acknowledges the support of a fellowship from La Caixa Foundation (ID 100010434) with fellowship code LCF/BQ/PI23/11970031 (P.I.: A. Escorza) and from the Fundación Occident and the Instituto de Astrofísica de Canarias under the  Visiting Researcher Programme 2022-2025 agreed between both institutions. SS aknowledges funding from the European Union under the grant ERC-2022-AdG, {\em "StarDance: the non-canonical evolution of stars in clusters"}, Grant Agreement 101093572, PI: E. Pancino. AM and MB acknowledge support from the project "LEGO – Reconstructing the building blocks of the Galaxy by chemical tagging" (P.I. A. Mucciarelli), granted by the Italian MUR through contract PRIN 2022LLP8TK\_001.
We acknowledge the support to this study by the INAF Mini Grant 2023 (Ob.Fu. 1.05.23.04.02 – CUP C33C23000960005) CHAM – Chemo-dynamics of the Accreted Halo of the Milky Way (P.I.: M. Bellazzini).
CG acknowledges support from the Agencia Estatal de Investigación del Ministerio de Ciencia e Innovación (AEI-MCINN) under grant "At the forefront of Galactic Archaeology: evolution of the luminous and dark matter components of the Milky Way and Local Group dwarf galaxies in the Gaia era" with reference PID2020-118778GB-I00/10.13039/501100011033. CG also acknowledge support from the Severo Ochoa program through CEX2019-000920-S.
ED acknowledges financial support from the INAF Data Analysis Research Grant (ref. E. Dalessandro) of the Bando Astrofisica Fondamentale 2024.
TRL acknowledges support from Juan de la Cierva fellowship (IJC2020-043742-I), financed by MCIN/AEI/10.13039/501100011033.
MM acknowledges support from the Agencia Estatal de Investigaci\'on del Ministerio de Ciencia e Innovaci\'on (MCIN/AEI) under the grant "RR Lyrae stars, a lighthouse to distant galaxies and early galaxy evolution" and the European Regional Development Fun (ERDF) with reference PID2021-127042OB-I00.
Co-funded by the European Union (ERC-2022-AdG, "StarDance: the non-canonical evolution of stars in clusters", Grant Agreement 101093572, PI: E. Pancino). Views and opinions expressed are however those of the author(s) only and do not necessarily reflect those of the European Union or the European Research Council. Neither the European Union nor the granting authority can be held responsible for them.
This paper is supported by the Italian Research Center on High Performance Computing Big Data and Quantum Computing (ICSC), project funded by European Union - NextGenerationEU - and National Recovery and Resilience Plan (NRRP) - Mission 4 Component 2 within the activities of Spoke 3 (Astrophysics and Cosmos Observations).
This work is part of the project Cosmic-Lab at the Physics and Astronomy Department “A. Righi” of the Bologna University (\url{http:// www.cosmic-lab.eu/ Cosmic-Lab/Home.html}).

Based on observations with the NASA/ESA HST, obtained
at the Space Telescope Science Institute, which is operated by
AURA, Inc., under NASA contract NAS 5-26555. Support for Program number GO-17435 was provided through grants from STScI under NASA contract NAS5-26555. This
research made use of emcee \citep{foreman-mackey2013emcee}.
This work has made use of data from the European Space Agency (ESA) mission
{\it Gaia}\ (\url{https://www.cosmos.esa.int/gaia}), processed by the {\it Gaia}\
Data Processing and Analysis Consortium (DPAC,
\url{https://www.cosmos.esa.int/web/gaia/dpac/consortium}). Funding for the DPAC
has been provided by national institutions, in particular the institutions
participating in the {\it Gaia}\ Multilateral Agreement.
This project has received funding from the European Research Council (ERC) under the European Union’s Horizon 2020 research and innovation programme (grant agreement No. 804240) for S.S. and Á.S. M.M. acknowledges support from the Agencia Estatal de Investigaci\'on del Ministerio de Ciencia e Innovaci\'on (MCIN/AEI) under the grant "RR Lyrae stars, a lighthouse to distant galaxies and early galaxy evolution" and the European Regional Development Fun (ERDF) with reference PID2021-127042OB-I00, and from the Spanish Ministry of Science and Innovation (MICINN) through the Spanish State Research Agency, under Severo Ochoa Programe 2020-2023 (CEX2019-000920-S).
. P.J. acknowledges support from the Swiss National Science Foundation.
\end{acknowledgements}

\bibliographystyle{aa}
\bibliography{aa54007-25corr.bib}

\begin{thebibliography}{110}
\expandafter\ifx\csname natexlab\endcsname\relax\def\natexlab#1{#1}\fi

\bibitem[{{Aguado-Agelet} {et~al.}(2025){Aguado-Agelet}, {Massari}, {Monelli}, {Cassisi}, {Gallart}, {Ceccarelli}, {Gonz{\'a}lez-Koda}, {Ruiz-Lara}, {Pancino}, {Saracino}, \& {Salaris}}]{aguado-agelet25}
{Aguado-Agelet}, F., {Massari}, D., {Monelli}, M., {et~al.} 2025, \aap~submitted, arXiv:2502.20436

\bibitem[{{Alvarez-Baena} {et~al.}(2024){Alvarez-Baena}, {Carrera}, {Thompson}, {Balaguer-Nu{\~n}ez}, {Bragaglia}, {Jordi}, {Silva-Villa}, \& {Vallenari}}]{alvarez24}
{Alvarez-Baena}, N., {Carrera}, R., {Thompson}, H., {et~al.} 2024, \aap, 687, A101

\bibitem[{{Alvarez Garay} {et~al.}(2024){Alvarez Garay}, {Fanelli}, {Origlia}, {Pallanca}, {Mucciarelli}, {Chiappino}, {Crociati}, {Lanzoni}, {Ferraro}, {Rich}, \& {Dalessandro}}]{deimer24}
{Alvarez Garay}, D.~A., {Fanelli}, C., {Origlia}, L., {et~al.} 2024, \aap, 686, A198

\bibitem[{{Anderson}(2016)}]{2016wfc..rept...12A}
{Anderson}, J. 2016, {Empirical Models for the WFC3/IR PSF}, Space Telescope WFC Instrument Science Report

\bibitem[{{Anderson} \& {King}(2006)}]{2006acs..rept....1A}
{Anderson}, J. \& {King}, I.~R. 2006, {PSFs, Photometry, and Astronomy for the ACS/WFC}, Instrument Science Report ACS 2006-01

\bibitem[{{Anderson} \& {Ryon}(2018)}]{2018acs..rept....4A}
{Anderson}, J. \& {Ryon}, J.~E. 2018, {Improving the Pixel-Based CTE-correction Model for ACS/WFC}, Instrument Science Report ACS 2018-04, 37 pages

\bibitem[{{Bastian} {et~al.}(2010){Bastian}, {Covey}, \& {Meyer}}]{bastian10}
{Bastian}, N., {Covey}, K.~R., \& {Meyer}, M.~R. 2010, \araa, 48, 339

\bibitem[{{Baumgardt} \& {Hilker}(2018)}]{baumgardt18}
{Baumgardt}, H. \& {Hilker}, M. 2018, \mnras, 478, 1520

\bibitem[{{Baumgardt} {et~al.}(2020){Baumgardt}, {Sollima}, \& {Hilker}}]{baumgardt20}
{Baumgardt}, H., {Sollima}, A., \& {Hilker}, M. 2020, \pasa, 37, e046

\bibitem[{{Baumgardt} \& {Vasiliev}(2021)}]{baumgardt21}
{Baumgardt}, H. \& {Vasiliev}, E. 2021, \mnras, 505, 5957

\bibitem[{{Bedin} {et~al.}(2004){Bedin}, {Piotto}, {Anderson}, {Cassisi}, {King}, {Momany}, \& {Carraro}}]{bedin04}
{Bedin}, L.~R., {Piotto}, G., {Anderson}, J., {et~al.} 2004, \apjl, 605, L125

\bibitem[{{Bedin} {et~al.}(2023){Bedin}, {Salaris}, {Anderson}, {Scalco}, {Nardiello}, {Vesperini}, {Richer}, {Burgasser}, {Griggio}, {Gerasimov}, {Apai}, {Bellini}, {Libralato}, {Bergeron}, {Rich}, \& {Grazian}}]{bedin23}
{Bedin}, L.~R., {Salaris}, M., {Anderson}, J., {et~al.} 2023, \mnras, 518, 3722

\bibitem[{{Bellini} {et~al.}(2011){Bellini}, {Anderson}, \& {Bedin}}]{2011PASP..123..622B}
{Bellini}, A., {Anderson}, J., \& {Bedin}, L.~R. 2011, \pasp, 123, 622

\bibitem[{{Bellini} {et~al.}(2017){Bellini}, {Anderson}, {Bedin}, {King}, {van der Marel}, {Piotto}, \& {Cool}}]{2017BelliniwCenI}
{Bellini}, A., {Anderson}, J., {Bedin}, L.~R., {et~al.} 2017, \apj, 842, 6

\bibitem[{{Bellini} \& {Bedin}(2009)}]{2009PASP..121.1419B}
{Bellini}, A. \& {Bedin}, L.~R. 2009, \pasp, 121, 1419

\bibitem[{{Bellini} {et~al.}(2013){Bellini}, {Piotto}, {Milone}, {King}, {Renzini}, {Cassisi}, {Anderson}, {Bedin}, {Nardiello}, {Pietrinferni}, \& {Sarajedini}}]{bellini13}
{Bellini}, A., {Piotto}, G., {Milone}, A.~P., {et~al.} 2013, \apj, 765, 32

\bibitem[{{Belokurov} {et~al.}(2018){Belokurov}, {Erkal}, {Evans}, {Koposov}, \& {Deason}}]{belokurov2018}
{Belokurov}, V., {Erkal}, D., {Evans}, N.~W., {Koposov}, S.~E., \& {Deason}, A.~J. 2018, \mnras, 478, 611

\bibitem[{{Belokurov} \& {Kravtsov}(2024)}]{belokurov24}
{Belokurov}, V. \& {Kravtsov}, A. 2024, \mnras, 528, 3198

\bibitem[{{Bica} {et~al.}(1999){Bica}, {Ortolani}, \& {Barbuy}}]{bica99}
{Bica}, E., {Ortolani}, S., \& {Barbuy}, B. 1999, \aaps, 136, 363

\bibitem[{{Bica} {et~al.}(2024){Bica}, {Ortolani}, {Barbuy}, \& {Oliveira}}]{bica24}
{Bica}, E., {Ortolani}, S., {Barbuy}, B., \& {Oliveira}, R.~A.~P. 2024, \aap, 687, A201

\bibitem[{{Bonatto} \& {Bica}(2008)}]{bonatto08}
{Bonatto}, C. \& {Bica}, E. 2008, \aap, 479, 741

\bibitem[{{Brown} {et~al.}(2016){Brown}, {Cassisi}, {D'Antona}, {Salaris}, {Milone}, {Dalessandro}, {Piotto}, {Renzini}, {Sweigart}, {Bellini}, {Ortolani}, {Sarajedini}, {Aparicio}, {Bedin}, {Anderson}, {Pietrinferni}, \& {Nardiello}}]{brown16}
{Brown}, T.~M., {Cassisi}, S., {D'Antona}, F., {et~al.} 2016, \apj, 822, 44

\bibitem[{{Cadelano} {et~al.}(2020){Cadelano}, {Dalessandro}, {Webb}, {Vesperini}, {Lattanzio}, {Beccari}, {Gomez}, \& {Monaco}}]{cadelano20}
{Cadelano}, M., {Dalessandro}, E., {Webb}, J.~J., {et~al.} 2020, \mnras, 499, 2390

\bibitem[{{Callingham} {et~al.}(2022){Callingham}, {Cautun}, {Deason}, {Frenk}, {Grand}, \& {Marinacci}}]{callingham22}
{Callingham}, T.~M., {Cautun}, M., {Deason}, A.~J., {et~al.} 2022, \mnras, 513, 4107

\bibitem[{{Cardelli} {et~al.}(1989){Cardelli}, {Clayton}, \& {Mathis}}]{cardelli89}
{Cardelli}, J.~A., {Clayton}, G.~C., \& {Mathis}, J.~S. 1989, \apj, 345, 245

\bibitem[{{Carretta} \& {Bragaglia}(2023)}]{carretta23}
{Carretta}, E. \& {Bragaglia}, A. 2023, \aap, 677, A73

\bibitem[{{Cassisi} {et~al.}(2004){Cassisi}, {Salaris}, {Castelli}, \& {Pietrinferni}}]{cassisi04}
{Cassisi}, S., {Salaris}, M., {Castelli}, F., \& {Pietrinferni}, A. 2004, \apj, 616, 498

\bibitem[{{Ceccarelli} {et~al.}(2025){Ceccarelli}, {Massari}, {Aguado-Agelet}, {Mucciarelli}, {Cassisi}, {Monelli}, {Pancino}, {Salaris}, \& {Saracino}}]{ceccarelli25}
{Ceccarelli}, E., {Massari}, D., {Aguado-Agelet}, F., {et~al.} 2025, \aap~submitted, arXiv:2503.02939

\bibitem[{{Ceccarelli} {et~al.}(2024{\natexlab{a}}){Ceccarelli}, {Massari}, {Mucciarelli}, {Bellazzini}, {Nunnari}, {Cusano}, {Lardo}, {Romano}, {Ilyin}, \& {Stokholm}}]{ceccarelli24b}
{Ceccarelli}, E., {Massari}, D., {Mucciarelli}, A., {et~al.} 2024{\natexlab{a}}, \aap, 684, A37

\bibitem[{{Ceccarelli} {et~al.}(2024{\natexlab{b}}){Ceccarelli}, {Mucciarelli}, {Massari}, {Bellazzini}, \& {Matsuno}}]{ceccarelli24a}
{Ceccarelli}, E., {Mucciarelli}, A., {Massari}, D., {Bellazzini}, M., \& {Matsuno}, T. 2024{\natexlab{b}}, \aap, 691, A226

\bibitem[{{Chaboyer}(1995)}]{chaboyer95}
{Chaboyer}, B. 1995, \apjl, 444, L9

\bibitem[{{Chen} \& {Gnedin}(2024)}]{chen24}
{Chen}, Y. \& {Gnedin}, O.~Y. 2024, The Open Journal of Astrophysics, 7, 23

\bibitem[{{Cordoni} {et~al.}(2023){Cordoni}, {Milone}, {Marino}, {Vesperini}, {Dondoglio}, {Legnardi}, {Mohandasan}, {Carlos}, {Lagioia}, {Jang}, \& {Ziliotto}}]{cordoni23}
{Cordoni}, G., {Milone}, A.~P., {Marino}, A.~F., {et~al.} 2023, \aap, 672, A29

\bibitem[{{Cornish} {et~al.}(2006){Cornish}, {Phelps}, {Briley}, \& {Friel}}]{cornish06}
{Cornish}, A. S.~M., {Phelps}, R.~L., {Briley}, M.~M., \& {Friel}, E.~D. 2006, \aj, 131, 2543

\bibitem[{{Crociati} {et~al.}(2023){Crociati}, {Valenti}, {Ferraro}, {Pallanca}, {Lanzoni}, {Cadelano}, {Fanelli}, {Origlia}, {Leanza}, {Dalessandro}, {Mucciarelli}, \& {Rich}}]{crociati23}
{Crociati}, C., {Valenti}, E., {Ferraro}, F.~R., {et~al.} 2023, \apj, 951, 17

\bibitem[{{Dalessandro} {et~al.}(2022){Dalessandro}, {Crociati}, {Cignoni}, {Ferraro}, {Lanzoni}, {Origlia}, {Pallanca}, {Rich}, {Saracino}, \& {Valenti}}]{dalessandro22}
{Dalessandro}, E., {Crociati}, C., {Cignoni}, M., {et~al.} 2022, \apj, 940, 170

\bibitem[{{De Marchi} {et~al.}(2010){De Marchi}, {Paresce}, \& {Portegies Zwart}}]{demarchi10}
{De Marchi}, G., {Paresce}, F., \& {Portegies Zwart}, S. 2010, \apj, 718, 105

\bibitem[{{del Pino} {et~al.}(2022){del Pino}, {Libralato}, {van der Marel}, {Bennet}, {Fardal}, {Anderson}, {Bellini}, {Tony Sohn}, \& {Watkins}}]{delpino22}
{del Pino}, A., {Libralato}, M., {van der Marel}, R.~P., {et~al.} 2022, \apj, 933, 76

\bibitem[{{Dotter} {et~al.}(2010){Dotter}, {Sarajedini}, {Anderson}, {Aparicio}, {Bedin}, {Chaboyer}, {Majewski}, {Mar{\'\i}n-Franch}, {Milone}, {Paust}, {Piotto}, {Reid}, {Rosenberg}, \& {Siegel}}]{dotter10}
{Dotter}, A., {Sarajedini}, A., {Anderson}, J., {et~al.} 2010, \apj, 708, 698

\bibitem[{{Fall} \& {Rees}(1985)}]{fall85}
{Fall}, S.~M. \& {Rees}, M.~J. 1985, \apj, 298, 18

\bibitem[{{Fanelli} {et~al.}(2024){Fanelli}, {Origlia}, {Rich}, {Ferraro}, {Alvarez Garay}, {Chiappino}, {Lanzoni}, {Pallanca}, {Crociati}, \& {Dalessandro}}]{fanelli24}
{Fanelli}, C., {Origlia}, L., {Rich}, R.~M., {et~al.} 2024, \aap, 690, A139

\bibitem[{{Ferraro} {et~al.}(2009){Ferraro}, {Dalessandro}, {Mucciarelli}, {Beccari}, {Rich}, {Origlia}, {Lanzoni}, {Rood}, {Valenti}, {Bellazzini}, {Ransom}, \& {Cocozza}}]{ferraro09}
{Ferraro}, F.~R., {Dalessandro}, E., {Mucciarelli}, A., {et~al.} 2009, \nat, 462, 483

\bibitem[{{Ferraro} {et~al.}(2012){Ferraro}, {Lanzoni}, {Dalessandro}, {Beccari}, {Pasquato}, {Miocchi}, {Rood}, {Sigurdsson}, {Sills}, {Vesperini}, {Mapelli}, {Contreras}, {Sanna}, \& {Mucciarelli}}]{ferraro12}
{Ferraro}, F.~R., {Lanzoni}, B., {Dalessandro}, E., {et~al.} 2012, \nat, 492, 393

\bibitem[{{Ferraro} {et~al.}(2016){Ferraro}, {Massari}, {Dalessandro}, {Lanzoni}, {Origlia}, {Rich}, \& {Mucciarelli}}]{ferraro16}
{Ferraro}, F.~R., {Massari}, D., {Dalessandro}, E., {et~al.} 2016, \apj, 828, 75

\bibitem[{{Ferraro} {et~al.}(2021){Ferraro}, {Pallanca}, {Lanzoni}, {Crociati}, {Dalessandro}, {Origlia}, {Rich}, {Saracino}, {Mucciarelli}, {Valenti}, {Geisler}, {Mauro}, {Villanova}, {Moni Bidin}, \& {Beccari}}]{ferraro21}
{Ferraro}, F.~R., {Pallanca}, C., {Lanzoni}, B., {et~al.} 2021, Nature Astronomy, 5, 311

\bibitem[{{Forbes}(2020)}]{forbes20}
{Forbes}, D.~A. 2020, \mnras, 493, 847

\bibitem[{{Forbes} \& {Bridges}(2010)}]{forbes2010}
{Forbes}, D.~A. \& {Bridges}, T. 2010, \mnras, 404, 1203

\bibitem[{Foreman-Mackey {et~al.}(2013)Foreman-Mackey, Hogg, Lang, \& Goodman}]{foreman-mackey2013emcee}
Foreman-Mackey, D., Hogg, D., Lang, D., \& Goodman, J. 2013, Publications of the Astronomical Society of the Pacific, 125, 306

\bibitem[{{Gaia Collaboration} {et~al.}(2016){Gaia Collaboration}, {Prusti}, {de Bruijne}, {Brown}, {Vallenari}, {Babusiaux}, {Bailer-Jones}, {Bastian}, {Biermann}, {Evans}, {Eyer}, {Jansen}, {Jordi}, {Klioner}, {Lammers}, {Lindegren}, {Luri}, {Mignard}, {Milligan}, {Panem}, {Poinsignon}, {Pourbaix}, {Randich}, {Sarri}, {Sartoretti}, {Siddiqui}, {Soubiran}, {Valette}, {van Leeuwen}, {Walton}, {Aerts}, {Arenou}, {Cropper}, {Drimmel}, {H{\o}g}, {Katz}, {Lattanzi}, {O'Mullane}, {Grebel}, {Holland}, {Huc}, {Passot}, {Bramante}, {Cacciari}, {Casta{\~n}eda}, {Chaoul}, {Cheek}, {De Angeli}, {Fabricius}, {Guerra}, {Hern{\'a}ndez}, {Jean-Antoine-Piccolo}, {Masana}, {Messineo}, {Mowlavi}, {Nienartowicz}, {Ord{\'o}{\~n}ez-Blanco}, {Panuzzo}, {Portell}, {Richards}, {Riello}, {Seabroke}, {Tanga}, {Th{\'e}venin}, {Torra}, {Els}, {Gracia-Abril}, {Comoretto}, {Garcia-Reinaldos}, {Lock}, {Mercier}, {Altmann}, {Andrae}, {Astraatmadja}, {Bellas-Velidis}, {Benson}, {Berthier}, {Blomme}, {Busso}, {Carry}, {Cellino}, {Clementini},
  {Cowell}, {Creevey}, {Cuypers}, {Davidson}, {De Ridder}, {de Torres}, {Delchambre}, {Dell'Oro}, {Ducourant}, {Fr{\'e}mat}, {Garc{\'\i}a-Torres}, {Gosset}, {Halbwachs}, {Hambly}, {Harrison}, {Hauser}, {Hestroffer}, {Hodgkin}, {Huckle}, {Hutton}, {Jasniewicz}, {Jordan}, {Kontizas}, {Korn}, {Lanzafame}, {Manteiga}, {Moitinho}, {Muinonen}, {Osinde}, {Pancino}, {Pauwels}, {Petit}, {Recio-Blanco}, {Robin}, {Sarro}, {Siopis}, {Smith}, {Smith}, {Sozzetti}, {Thuillot}, {van Reeven}, {Viala}, {Abbas}, {Abreu Aramburu}, {Accart}, {Aguado}, {Allan}, {Allasia}, {Altavilla}, {{\'A}lvarez}, {Alves}, {Anderson}, {Andrei}, {Anglada Varela}, {Antiche}, {Antoja}, {Ant{\'o}n}, {Arcay}, {Atzei}, {Ayache}, {Bach}, {Baker}, {Balaguer-N{\'u}{\~n}ez}, {Barache}, {Barata}, {Barbier}, {Barblan}, {Baroni}, {Barrado y Navascu{\'e}s}, {Barros}, {Barstow}, {Becciani}, {Bellazzini}, {Bellei}, {Bello Garc{\'\i}a}, {Belokurov}, {Bendjoya}, {Berihuete}, {Bianchi}, {Bienaym{\'e}}, {Billebaud}, {Blagorodnova}, {Blanco-Cuaresma}, {Boch},
  {Bombrun}, {Borrachero}, {Bouquillon}, {Bourda}, {Bouy}, {Bragaglia}, {Breddels}, {Brouillet}, {Br{\"u}semeister}, {Bucciarelli}, {Budnik}, {Burgess}, {Burgon}, {Burlacu}, {Busonero}, {Buzzi}, {Caffau}, {Cambras}, {Campbell}, {Cancelliere}, {Cantat-Gaudin}, {Carlucci}, {Carrasco}, {Castellani}, {Charlot}, {Charnas}, {Charvet}, {Chassat}, {Chiavassa}, {Clotet}, {Cocozza}, {Collins}, {Collins}, {Costigan}, {Crifo}, {Cross}, {Crosta}, {Crowley}, {Dafonte}, {Damerdji}, {Dapergolas}, {David}, {David}, {De Cat}, {de Felice}, {de Laverny}, {De Luise}, {De March}, {de Martino}, {de Souza}, {Debosscher}, {del Pozo}, {Delbo}, {Delgado}, {Delgado}, {di Marco}, {Di Matteo}, {Diakite}, {Distefano}, {Dolding}, {Dos Anjos}, {Drazinos}, {Dur{\'a}n}, {Dzigan}, {Ecale}, {Edvardsson}, {Enke}, {Erdmann}, {Escolar}, {Espina}, {Evans}, {Eynard Bontemps}, {Fabre}, {Fabrizio}, {Faigler}, {Falc{\~a}o}, {Farr{\`a}s Casas}, {Faye}, {Federici}, {Fedorets}, {Fern{\'a}ndez-Hern{\'a}ndez}, {Fernique}, {Fienga}, {Figueras}, {Filippi},
  {Findeisen}, {Fonti}, {Fouesneau}, {Fraile}, {Fraser}, {Fuchs}, {Furnell}, {Gai}, {Galleti}, {Galluccio}, {Garabato}, {Garc{\'\i}a-Sedano}, {Gar{\'e}}, {Garofalo}, {Garralda}, {Gavras}, {Gerssen}, {Geyer}, {Gilmore}, {Girona}, {Giuffrida}, {Gomes}, {Gonz{\'a}lez-Marcos}, {Gonz{\'a}lez-N{\'u}{\~n}ez}, {Gonz{\'a}lez-Vidal}, {Granvik}, {Guerrier}, {Guillout}, {Guiraud}, {G{\'u}rpide}, {Guti{\'e}rrez-S{\'a}nchez}, {Guy}, {Haigron}, {Hatzidimitriou}, {Haywood}, {Heiter}, {Helmi}, {Hobbs}, {Hofmann}, {Holl}, {Holland}, {Hunt}, {Hypki}, {Icardi}, {Irwin}, {Jevardat de Fombelle}, {Jofr{\'e}}, {Jonker}, {Jorissen}, {Julbe}, {Karampelas}, {Kochoska}, {Kohley}, {Kolenberg}, {Kontizas}, {Koposov}, {Kordopatis}, {Koubsky}, {Kowalczyk}, {Krone-Martins}, {Kudryashova}, {Kull}, {Bachchan}, {Lacoste-Seris}, {Lanza}, {Lavigne}, {Le Poncin-Lafitte}, {Lebreton}, {Lebzelter}, {Leccia}, {Leclerc}, {Lecoeur-Taibi}, {Lemaitre}, {Lenhardt}, {Leroux}, {Liao}, {Licata}, {Lindstr{\o}m}, {Lister}, {Livanou}, {Lobel}, {L{\"o}ffler},
  {L{\'o}pez}, {Lopez-Lozano}, {Lorenz}, {Loureiro}, {MacDonald}, {Magalh{\~a}es Fernandes}, {Managau}, {Mann}, {Mantelet}, {Marchal}, {Marchant}, {Marconi}, {Marie}, {Marinoni}, {Marrese}, {Marschalk{\'o}}, {Marshall}, {Mart{\'\i}n-Fleitas}, {Martino}, {Mary}, {Matijevi{\v{c}}}, {Mazeh}, {McMillan}, {Messina}, {Mestre}, {Michalik}, {Millar}, {Miranda}, {Molina}, {Molinaro}, {Molinaro}, {Moln{\'a}r}, {Moniez}, {Montegriffo}, {Monteiro}, {Mor}, {Mora}, {Morbidelli}, {Morel}, {Morgenthaler}, {Morley}, {Morris}, {Mulone}, {Muraveva}, {Musella}, {Narbonne}, {Nelemans}, {Nicastro}, {Noval}, {Ord{\'e}novic}, {Ordieres-Mer{\'e}}, {Osborne}, {Pagani}, {Pagano}, {Pailler}, {Palacin}, {Palaversa}, {Parsons}, {Paulsen}, {Pecoraro}, {Pedrosa}, {Pentik{\"a}inen}, {Pereira}, {Pichon}, {Piersimoni}, {Pineau}, {Plachy}, {Plum}, {Poujoulet}, {Pr{\v{s}}a}, {Pulone}, {Ragaini}, {Rago}, {Rambaux}, {Ramos-Lerate}, {Ranalli}, {Rauw}, {Read}, {Regibo}, {Renk}, {Reyl{\'e}}, {Ribeiro}, {Rimoldini}, {Ripepi}, {Riva}, {Rixon},
  {Roelens}, {Romero-G{\'o}mez}, {Rowell}, {Royer}, {Rudolph}, {Ruiz-Dern}, {Sadowski}, {Sagrist{\`a} Sell{\'e}s}, {Sahlmann}, {Salgado}, {Salguero}, {Sarasso}, {Savietto}, {Schnorhk}, {Schultheis}, {Sciacca}, {Segol}, {Segovia}, {Segransan}, {Serpell}, {Shih}, {Smareglia}, {Smart}, {Smith}, {Solano}, {Solitro}, {Sordo}, {Soria Nieto}, {Souchay}, {Spagna}, {Spoto}, {Stampa}, {Steele}, {Steidelm{\"u}ller}, {Stephenson}, {Stoev}, {Suess}, {S{\"u}veges}, {Surdej}, {Szabados}, {Szegedi-Elek}, {Tapiador}, {Taris}, {Tauran}, {Taylor}, {Teixeira}, {Terrett}, {Tingley}, {Trager}, {Turon}, {Ulla}, {Utrilla}, {Valentini}, {van Elteren}, {Van Hemelryck}, {van Leeuwen}, {Varadi}, {Vecchiato}, {Veljanoski}, {Via}, {Vicente}, {Vogt}, {Voss}, {Votruba}, {Voutsinas}, {Walmsley}, {Weiler}, {Weingrill}, {Werner}, {Wevers}, {Whitehead}, {Wyrzykowski}, {Yoldas}, {{\v{Z}}erjal}, {Zucker}, {Zurbach}, {Zwitter}, {Alecu}, {Allen}, {Allende Prieto}, {Amorim}, {Anglada-Escud{\'e}}, {Arsenijevic}, {Azaz}, {Balm}, {Beck}, {Bernstein},
  {Bigot}, {Bijaoui}, {Blasco}, {Bonfigli}, {Bono}, {Boudreault}, {Bressan}, {Brown}, {Brunet}, {Bunclark}, {Buonanno}, {Butkevich}, {Carret}, {Carrion}, {Chemin}, {Ch{\'e}reau}, {Corcione}, {Darmigny}, {de Boer}, {de Teodoro}, {de Zeeuw}, {Delle Luche}, {Domingues}, {Dubath}, {Fodor}, {Fr{\'e}zouls}, {Fries}, {Fustes}, {Fyfe}, {Gallardo}, {Gallegos}, {Gardiol}, {Gebran}, {Gomboc}, {G{\'o}mez}, {Grux}, {Gueguen}, {Heyrovsky}, {Hoar}, {Iannicola}, {Isasi Parache}, {Janotto}, {Joliet}, {Jonckheere}, {Keil}, {Kim}, {Klagyivik}, {Klar}, {Knude}, {Kochukhov}, {Kolka}, {Kos}, {Kutka}, {Lainey}, {LeBouquin}, {Liu}, {Loreggia}, {Makarov}, {Marseille}, {Martayan}, {Martinez-Rubi}, {Massart}, {Meynadier}, {Mignot}, {Munari}, {Nguyen}, {Nordlander}, {Ocvirk}, {O'Flaherty}, {Olias Sanz}, {Ortiz}, {Osorio}, {Oszkiewicz}, {Ouzounis}, {Palmer}, {Park}, {Pasquato}, {Peltzer}, {Peralta}, {P{\'e}turaud}, {Pieniluoma}, {Pigozzi}, {Poels}, {Prat}, {Prod'homme}, {Raison}, {Rebordao}, {Risquez}, {Rocca-Volmerange}, {Rosen},
  {Ruiz-Fuertes}, {Russo}, {Sembay}, {Serraller Vizcaino}, {Short}, {Siebert}, {Silva}, {Sinachopoulos}, {Slezak}, {Soffel}, {Sosnowska}, {Strai{\v{z}}ys}, {ter Linden}, {Terrell}, {Theil}, {Tiede}, {Troisi}, {Tsalmantza}, {Tur}, {Vaccari}, {Vachier}, {Valles}, {Van Hamme}, {Veltz}, {Virtanen}, {Wallut}, {Wichmann}, {Wilkinson}, {Ziaeepour}, \& {Zschocke}}]{gaia}
{Gaia Collaboration}, {Prusti}, T., {de Bruijne}, J.~H.~J., {et~al.} 2016, \aap, 595, A1

\bibitem[{{Gaia Collaboration} {et~al.}(2023){Gaia Collaboration}, {Vallenari}, {Brown}, {Prusti}, {de Bruijne}, {Arenou}, {Babusiaux}, {Biermann}, {Creevey}, {Ducourant}, {Evans}, {Eyer}, {Guerra}, {Hutton}, {Jordi}, {Klioner}, {Lammers}, {Lindegren}, {Luri}, {Mignard}, {Panem}, {Pourbaix}, {Randich}, {Sartoretti}, {Soubiran}, {Tanga}, {Walton}, {Bailer-Jones}, {Bastian}, {Drimmel}, {Jansen}, {Katz}, {Lattanzi}, {van Leeuwen}, {Bakker}, {Cacciari}, {Casta{\~n}eda}, {De Angeli}, {Fabricius}, {Fouesneau}, {Fr{\'e}mat}, {Galluccio}, {Guerrier}, {Heiter}, {Masana}, {Messineo}, {Mowlavi}, {Nicolas}, {Nienartowicz}, {Pailler}, {Panuzzo}, {Riclet}, {Roux}, {Seabroke}, {Sordo}, {Th{\'e}venin}, {Gracia-Abril}, {Portell}, {Teyssier}, {Altmann}, {Andrae}, {Audard}, {Bellas-Velidis}, {Benson}, {Berthier}, {Blomme}, {Burgess}, {Busonero}, {Busso}, {C{\'a}novas}, {Carry}, {Cellino}, {Cheek}, {Clementini}, {Damerdji}, {Davidson}, {de Teodoro}, {Nu{\~n}ez Campos}, {Delchambre}, {Dell'Oro}, {Esquej},
  {Fern{\'a}ndez-Hern{\'a}ndez}, {Fraile}, {Garabato}, {Garc{\'\i}a-Lario}, {Gosset}, {Haigron}, {Halbwachs}, {Hambly}, {Harrison}, {Hern{\'a}ndez}, {Hestroffer}, {Hodgkin}, {Holl}, {Jan{\ss}en}, {Jevardat de Fombelle}, {Jordan}, {Krone-Martins}, {Lanzafame}, {L{\"o}ffler}, {Marchal}, {Marrese}, {Moitinho}, {Muinonen}, {Osborne}, {Pancino}, {Pauwels}, {Recio-Blanco}, {Reyl{\'e}}, {Riello}, {Rimoldini}, {Roegiers}, {Rybizki}, {Sarro}, {Siopis}, {Smith}, {Sozzetti}, {Utrilla}, {van Leeuwen}, {Abbas}, {{\'A}brah{\'a}m}, {Abreu Aramburu}, {Aerts}, {Aguado}, {Ajaj}, {Aldea-Montero}, {Altavilla}, {{\'A}lvarez}, {Alves}, {Anders}, {Anderson}, {Anglada Varela}, {Antoja}, {Baines}, {Baker}, {Balaguer-N{\'u}{\~n}ez}, {Balbinot}, {Balog}, {Barache}, {Barbato}, {Barros}, {Barstow}, {Bartolom{\'e}}, {Bassilana}, {Bauchet}, {Becciani}, {Bellazzini}, {Berihuete}, {Bernet}, {Bertone}, {Bianchi}, {Binnenfeld}, {Blanco-Cuaresma}, {Blazere}, {Boch}, {Bombrun}, {Bossini}, {Bouquillon}, {Bragaglia}, {Bramante}, {Breedt},
  {Bressan}, {Brouillet}, {Brugaletta}, {Bucciarelli}, {Burlacu}, {Butkevich}, {Buzzi}, {Caffau}, {Cancelliere}, {Cantat-Gaudin}, {Carballo}, {Carlucci}, {Carnerero}, {Carrasco}, {Casamiquela}, {Castellani}, {Castro-Ginard}, {Chaoul}, {Charlot}, {Chemin}, {Chiaramida}, {Chiavassa}, {Chornay}, {Comoretto}, {Contursi}, {Cooper}, {Cornez}, {Cowell}, {Crifo}, {Cropper}, {Crosta}, {Crowley}, {Dafonte}, {Dapergolas}, {David}, {David}, {de Laverny}, {De Luise}, \& {De March}}]{gaiadr3}
{Gaia Collaboration}, {Vallenari}, A., {Brown}, A.~G.~A., {et~al.} 2023, \aap, 674, A1

\bibitem[{{Garro} {et~al.}(2023){Garro}, {Fern{\'a}ndez-Trincado}, {Minniti}, {Moya}, {Palma}, {Beers}, {Placco}, {Barbuy}, {Sneden}, {Alves-Brito}, {Dias}, {Af{\c{s}}ar}, {Frelijj}, \& {Lane}}]{garro23}
{Garro}, E.~R., {Fern{\'a}ndez-Trincado}, J.~G., {Minniti}, D., {et~al.} 2023, \aap, 669, A136

\bibitem[{{Garro} {et~al.}(2024){Garro}, {Minniti}, \& {Fern{\'a}ndez-Trincado}}]{garro24}
{Garro}, E.~R., {Minniti}, D., \& {Fern{\'a}ndez-Trincado}, J.~G. 2024, \aap, 687, A214

\bibitem[{{Girardi} {et~al.}(2008){Girardi}, {Dalcanton}, {Williams}, {de Jong}, {Gallart}, {Monelli}, {Groenewegen}, {Holtzman}, {Olsen}, {Seth}, {Weisz}, \& {ANGST/ANGRRR Collaboration}}]{girardi08}
{Girardi}, L., {Dalcanton}, J., {Williams}, B., {et~al.} 2008, \pasp, 120, 583

\bibitem[{{Gratton} {et~al.}(2004){Gratton}, {Sneden}, \& {Carretta}}]{gratton04}
{Gratton}, R., {Sneden}, C., \& {Carretta}, E. 2004, \araa, 42, 385

\bibitem[{{H{\"a}berle} {et~al.}(2021){H{\"a}berle}, {Libralato}, {Bellini}, {Watkins}, {Pott}, {Neumayer}, {van der Marel}, {Piotto}, \& {Nardiello}}]{haberle21}
{H{\"a}berle}, M., {Libralato}, M., {Bellini}, A., {et~al.} 2021, \mnras, 503, 1490

\bibitem[{{Harris}(1996)}]{harris96}
{Harris}, W.~E. 1996, \aj, 112, 1487

\bibitem[{{Helmi} {et~al.}(2018){Helmi}, {Babusiaux}, {Koppelman}, {Massari}, {Veljanoski}, \& {Brown}}]{helmi2018}
{Helmi}, A., {Babusiaux}, C., {Koppelman}, H.~H., {et~al.} 2018, \nat, 563, 85

\bibitem[{{H{\'e}nault-Brunet} {et~al.}(2019){H{\'e}nault-Brunet}, {Gieles}, {Sollima}, {Watkins}, {Zocchi}, {Claydon}, {Pancino}, \& {Baumgardt}}]{henault19}
{H{\'e}nault-Brunet}, V., {Gieles}, M., {Sollima}, A., {et~al.} 2019, \mnras, 483, 1400

\bibitem[{{Hidalgo} {et~al.}(2018){Hidalgo}, {Pietrinferni}, {Cassisi}, {Salaris}, {Mucciarelli}, {Savino}, {Aparicio}, {Silva Aguirre}, \& {Verma}}]{hidalgo18}
{Hidalgo}, S.~L., {Pietrinferni}, A., {Cassisi}, S., {et~al.} 2018, \apj, 856, 125

\bibitem[{{Horta} {et~al.}(2020){Horta}, {Schiavon}, {Mackereth}, {Beers}, {Fern{\'a}ndez-Trincado}, {Frinchaboy}, {Garc{\'\i}a-Hern{\'a}ndez}, {Geisler}, {Hasselquist}, {J{\"o}nsson}, {Lane}, {Majewski}, {M{\'e}sz{\'a}ros}, {Bidin}, {Nataf}, {Roman-Lopes}, {Nitschelm}, {Vargas-Gonz{\'a}lez}, \& {Zasowski}}]{horta20}
{Horta}, D., {Schiavon}, R.~P., {Mackereth}, J.~T., {et~al.} 2020, \mnras, 493, 3363

\bibitem[{{Hunt} \& {Reffert}(2024)}]{hunt_reffert24}
{Hunt}, E.~L. \& {Reffert}, S. 2024, \aap, 686, A42

\bibitem[{{Hurt} {et~al.}(2000){Hurt}, {Jarrett}, {Kirkpatrick}, {Cutri}, {Schneider}, {Skrutskie}, \& {van Driel}}]{hurt2000}
{Hurt}, R.~L., {Jarrett}, T.~H., {Kirkpatrick}, J.~D., {et~al.} 2000, \aj, 120, 1876

\bibitem[{{Ivanov} \& {Borissova}(2002)}]{ivanov02}
{Ivanov}, V.~D. \& {Borissova}, J. 2002, \aap, 390, 937

\bibitem[{{Just} {et~al.}(2023){Just}, {Piskunov}, {Klos}, {Kovaleva}, \& {Polyachenko}}]{just23}
{Just}, A., {Piskunov}, A.~E., {Klos}, J.~H., {Kovaleva}, D.~A., \& {Polyachenko}, E.~V. 2023, \aap, 672, A187

\bibitem[{{Kinman}(1959)}]{kinman59}
{Kinman}, T.~D. 1959, \mnras, 119, 538

\bibitem[{{Koch} {et~al.}(2017){Koch}, {Hansen}, \& {Kunder}}]{koch17}
{Koch}, A., {Hansen}, C.~J., \& {Kunder}, A. 2017, \aap, 604, A41

\bibitem[{{Koposov} {et~al.}(2007){Koposov}, {de Jong}, {Belokurov}, {Rix}, {Zucker}, {Evans}, {Gilmore}, {Irwin}, \& {Bell}}]{koposov07}
{Koposov}, S., {de Jong}, J.~T.~A., {Belokurov}, V., {et~al.} 2007, \apj, 669, 337

\bibitem[{{Kruijssen} {et~al.}(2019){Kruijssen}, {Pfeffer}, {Reina-Campos}, {Crain}, \& {Bastian}}]{kruijssen19}
{Kruijssen}, J.~M.~D., {Pfeffer}, J.~L., {Reina-Campos}, M., {Crain}, R.~A., \& {Bastian}, N. 2019, \mnras, 486, 3180

\bibitem[{{Lauberts} {et~al.}(1981){Lauberts}, {Holmberg}, {Schuster}, \& {West}}]{lauberts81}
{Lauberts}, A., {Holmberg}, E.~B., {Schuster}, H.~E., \& {West}, R.~M. 1981, \aaps, 43, 307

\bibitem[{{Leaman} {et~al.}(2013){Leaman}, {VandenBerg}, \& {Mendel}}]{leaman13}
{Leaman}, R., {VandenBerg}, D.~A., \& {Mendel}, J.~T. 2013, \mnras, 436, 122

\bibitem[{{Libralato} {et~al.}(2023){Libralato}, {Bellini}, {van der Marel}, {Anderson}, {Sohn}, {Watkins}, {Alderson}, {Allen}, {Clampin}, {Glidden}, {Goyal}, {Hoch}, {Huang}, {Kammerer}, {Lewis}, {Lin}, {Long}, {Louie}, {MacDonald}, {Mountain}, {Pe{\~n}a-Guerrero}, {Perrin}, {Pueyo}, {Rebollido}, {Rickman}, {Seager}, {Stevenson}, {Valenti}, {Valentine}, \& {Wakeford}}]{libralato23}
{Libralato}, M., {Bellini}, A., {van der Marel}, R.~P., {et~al.} 2023, \apj, 950, 101

\bibitem[{{Libralato} {et~al.}(2022){Libralato}, {Bellini}, {Vesperini}, {Piotto}, {Milone}, {van der Marel}, {Anderson}, {Aparicio}, {Barbuy}, {Bedin}, {Borsato}, {Cassisi}, {Dalessandro}, {Ferraro}, {King}, {Lanzoni}, {Nardiello}, {Ortolani}, {Sarajedini}, \& {Sohn}}]{libralato22}
{Libralato}, M., {Bellini}, A., {Vesperini}, E., {et~al.} 2022, \apj, 934, 150

\bibitem[{{Magrini} {et~al.}(2023){Magrini}, {Viscasillas V{\'a}zquez}, {Spina}, {Randich}, {Romano}, {Franciosini}, {Recio-Blanco}, {Nordlander}, {D'Orazi}, {Baratella}, {Smiljanic}, {Dantas}, {Pasquini}, {Spitoni}, {Casali}, {Van der Swaelmen}, {Bensby}, {Stonkute}, {Feltzing}, {Sacco}, {Bragaglia}, {Pancino}, {Heiter}, {Biazzo}, {Gilmore}, {Bergemann}, {Tautvai{\v{s}}ien{\.{e}}}, {Worley}, {Hourihane}, {Gonneau}, \& {Morbidelli}}]{magrini23}
{Magrini}, L., {Viscasillas V{\'a}zquez}, C., {Spina}, L., {et~al.} 2023, \aap, 669, A119

\bibitem[{{Malhan} {et~al.}(2022){Malhan}, {Ibata}, {Sharma}, {Famaey}, {Bellazzini}, {Carlberg}, {D'Souza}, {Yuan}, {Martin}, \& {Thomas}}]{malhan22}
{Malhan}, K., {Ibata}, R.~A., {Sharma}, S., {et~al.} 2022, \apj, 926, 107

\bibitem[{{Mar{\'\i}n-Franch} {et~al.}(2009){Mar{\'\i}n-Franch}, {Aparicio}, {Piotto}, {Rosenberg}, {Chaboyer}, {Sarajedini}, {Siegel}, {Anderson}, {Bedin}, {Dotter}, {Hempel}, {King}, {Majewski}, {Milone}, {Paust}, \& {Reid}}]{marin-franch09}
{Mar{\'\i}n-Franch}, A., {Aparicio}, A., {Piotto}, G., {et~al.} 2009, \apj, 694, 1498

\bibitem[{{Massari} {et~al.}(2023){Massari}, {Aguado-Agelet}, {Monelli}, {Cassisi}, {Pancino}, {Saracino}, {Gallart}, {Ruiz-Lara}, {Fern{\'a}ndez-Alvar}, {Surot}, {Stokholm}, {Salaris}, {Miglio}, \& {Ceccarelli}}]{massari23}
{Massari}, D., {Aguado-Agelet}, F., {Monelli}, M., {et~al.} 2023, \aap, 680, A20

\bibitem[{{Massari} {et~al.}(2018){Massari}, {Breddels}, {Helmi}, {Posti}, {Brown}, \& {Tolstoy}}]{massari18}
{Massari}, D., {Breddels}, M.~A., {Helmi}, A., {et~al.} 2018, Nature Astronomy, 2, 156

\bibitem[{{Massari} {et~al.}(2020){Massari}, {Helmi}, {Mucciarelli}, {Sales}, {Spina}, \& {Tolstoy}}]{massari20}
{Massari}, D., {Helmi}, A., {Mucciarelli}, A., {et~al.} 2020, \aap, 633, A36

\bibitem[{{Massari} {et~al.}(2019){Massari}, {Koppelman}, \& {Helmi}}]{massari19}
{Massari}, D., {Koppelman}, H.~H., \& {Helmi}, A. 2019, \aap, 630, L4

\bibitem[{{Massari} {et~al.}(2014){Massari}, {Mucciarelli}, {Ferraro}, {Origlia}, {Rich}, {Lanzoni}, {Dalessandro}, {Valenti}, {Ibata}, {Lovisi}, {Bellazzini}, \& {Reitzel}}]{massari14}
{Massari}, D., {Mucciarelli}, A., {Ferraro}, F.~R., {et~al.} 2014, \apj, 795, 22

\bibitem[{{Massari} {et~al.}(2017){Massari}, {Posti}, {Helmi}, {Fiorentino}, \& {Tolstoy}}]{massari17}
{Massari}, D., {Posti}, L., {Helmi}, A., {Fiorentino}, G., \& {Tolstoy}, E. 2017, \aap, 598, L9

\bibitem[{{McMillan}(2017)}]{mcmillan17}
{McMillan}, P.~J. 2017, \mnras, 465, 76

\bibitem[{{Milone} \& {Marino}(2022)}]{milone22}
{Milone}, A.~P. \& {Marino}, A.~F. 2022, Universe, 8, 359

\bibitem[{{Milone} {et~al.}(2012){Milone}, {Piotto}, {Bedin}, {Aparicio}, {Anderson}, {Sarajedini}, {Marino}, {Moretti}, {Davies}, {Chaboyer}, {Dotter}, {Hempel}, {Mar{\'\i}n-Franch}, {Majewski}, {Paust}, {Reid}, {Rosenberg}, \& {Siegel}}]{milone12}
{Milone}, A.~P., {Piotto}, G., {Bedin}, L.~R., {et~al.} 2012, \aap, 540, A16

\bibitem[{{Milone} {et~al.}(2017){Milone}, {Piotto}, {Renzini}, {Marino}, {Bedin}, {Vesperini}, {D'Antona}, {Nardiello}, {Anderson}, {King}, {Yong}, {Bellini}, {Aparicio}, {Barbuy}, {Brown}, {Cassisi}, {Ortolani}, {Salaris}, {Sarajedini}, \& {van der Marel}}]{milone17}
{Milone}, A.~P., {Piotto}, G., {Renzini}, A., {et~al.} 2017, \mnras, 464, 3636

\bibitem[{{Minniti} {et~al.}(1995){Minniti}, {Olszewski}, \& {Rieke}}]{minniti95}
{Minniti}, D., {Olszewski}, E.~W., \& {Rieke}, M. 1995, \aj, 110, 1686

\bibitem[{{Monty} {et~al.}(2023){Monty}, {Yong}, {Massari}, {McKenzie}, {Myeong}, {Buder}, {Karakas}, {Freeman}, {Marino}, {Belokurov}, \& {Evans}}]{monty23}
{Monty}, S., {Yong}, D., {Massari}, D., {et~al.} 2023, \mnras, 522, 4404

\bibitem[{{Nardiello} {et~al.}(2018){Nardiello}, {Libralato}, {Piotto}, {Anderson}, {Bellini}, {Aparicio}, {Bedin}, {Cassisi}, {Granata}, {King}, {Lucertini}, {Marino}, {Milone}, {Ortolani}, {Platais}, \& {van der Marel}}]{nardiello2018}
{Nardiello}, D., {Libralato}, M., {Piotto}, G., {et~al.} 2018, \mnras, 481, 3382

\bibitem[{{Origlia} {et~al.}(2013){Origlia}, {Massari}, {Rich}, {Mucciarelli}, {Ferraro}, {Dalessandro}, \& {Lanzoni}}]{origlia13}
{Origlia}, L., {Massari}, D., {Rich}, R.~M., {et~al.} 2013, \apjl, 779, L5

\bibitem[{{Origlia} {et~al.}(2011){Origlia}, {Rich}, {Ferraro}, {Lanzoni}, {Bellazzini}, {Dalessandro}, {Mucciarelli}, {Valenti}, \& {Beccari}}]{origlia11}
{Origlia}, L., {Rich}, R.~M., {Ferraro}, F.~R., {et~al.} 2011, \apjl, 726, L20

\bibitem[{{Pace} {et~al.}(2023){Pace}, {Koposov}, {Walker}, {Caldwell}, {Mateo}, {Olszewski}, {Roederer}, {Bailey}, {Belokurov}, {Kuehn}, {Li}, \& {Zucker}}]{pace23}
{Pace}, A.~B., {Koposov}, S.~E., {Walker}, M.~G., {et~al.} 2023, \mnras, 526, 1075

\bibitem[{{Palla} {et~al.}(2024){Palla}, {Magrini}, {Spitoni}, {Matteucci}, {Viscasillas V{\'a}zquez}, {Franchini}, {Molero}, \& {Randich}}]{palla24}
{Palla}, M., {Magrini}, L., {Spitoni}, E., {et~al.} 2024, \aap, 690, A334

\bibitem[{{Paust} {et~al.}(2010){Paust}, {Reid}, {Piotto}, {Aparicio}, {Anderson}, {Sarajedini}, {Bedin}, {Chaboyer}, {Dotter}, {Hempel}, {Majewski}, {Mar{\'\i}n-Franch}, {Milone}, {Rosenberg}, \& {Siegel}}]{paust10}
{Paust}, N. E.~Q., {Reid}, I.~N., {Piotto}, G., {et~al.} 2010, \aj, 139, 476

\bibitem[{{Pietrinferni} {et~al.}(2021){Pietrinferni}, {Hidalgo}, {Cassisi}, {Salaris}, {Savino}, {Mucciarelli}, {Verma}, {Silva Aguirre}, {Aparicio}, \& {Ferguson}}]{pietrinferni21}
{Pietrinferni}, A., {Hidalgo}, S., {Cassisi}, S., {et~al.} 2021, \apj, 908, 102

\bibitem[{{Piotto} {et~al.}(2015){Piotto}, {Milone}, {Bedin}, {Anderson}, {King}, {Marino}, {Nardiello}, {Aparicio}, {Barbuy}, {Bellini}, {Brown}, {Cassisi}, {Cool}, {Cunial}, {Dalessandro}, {D'Antona}, {Ferraro}, {Hidalgo}, {Lanzoni}, {Monelli}, {Ortolani}, {Renzini}, {Salaris}, {Sarajedini}, {van der Marel}, {Vesperini}, \& {Zoccali}}]{piotto15}
{Piotto}, G., {Milone}, A.~P., {Bedin}, L.~R., {et~al.} 2015, \aj, 149, 91

\bibitem[{{Recio-Blanco}(2018)}]{recio-blanco18}
{Recio-Blanco}, A. 2018, \aap, 620, A194

\bibitem[{{Salaris} {et~al.}(2002){Salaris}, {Cassisi}, \& {Weiss}}]{salaris02}
{Salaris}, M., {Cassisi}, S., \& {Weiss}, A. 2002, \pasp, 114, 375

\bibitem[{{Salaris} {et~al.}(1993){Salaris}, {Chieffi}, \& {Straniero}}]{salaris93}
{Salaris}, M., {Chieffi}, A., \& {Straniero}, O. 1993, \apj, 414, 580

\bibitem[{{Sarajedini} {et~al.}(2007){Sarajedini}, {Bedin}, {Chaboyer}, {Dotter}, {Siegel}, {Anderson}, {Aparicio}, {King}, {Majewski}, {Mar{\'\i}n-Franch}, {Piotto}, {Reid}, \& {Rosenberg}}]{sarajedini07}
{Sarajedini}, A., {Bedin}, L.~R., {Chaboyer}, B., {et~al.} 2007, \aj, 133, 1658

\bibitem[{{Schiavon} {et~al.}(2024){Schiavon}, {Phillips}, {Myers}, {Horta}, {Minniti}, {Allende Prieto}, {Anguiano}, {Beaton}, {Beers}, {Brownstein}, {Cohen}, {Fern{\'a}ndez-Trincado}, {Frinchaboy}, {J{\"o}nsson}, {Kisku}, {Lane}, {Majewski}, {Mason}, {M{\'e}sz{\'a}ros}, \& {Stringfellow}}]{schiavon24}
{Schiavon}, R.~P., {Phillips}, S.~G., {Myers}, N., {et~al.} 2024, \mnras, 528, 1393

\bibitem[{{Searle} \& {Zinn}(1978)}]{searle78}
{Searle}, L. \& {Zinn}, R. 1978, \apj, 225, 357

\bibitem[{{Simpson} {et~al.}(2017){Simpson}, {De Silva}, {Martell}, {Navin}, \& {Zucker}}]{simpson17}
{Simpson}, J.~D., {De Silva}, G., {Martell}, S.~L., {Navin}, C.~A., \& {Zucker}, D.~B. 2017, \mnras, 472, 2856

\bibitem[{{Skrutskie} {et~al.}(2006){Skrutskie}, {Cutri}, {Stiening}, {Weinberg}, {Schneider}, {Carpenter}, {Beichman}, {Capps}, {Chester}, {Elias}, {Huchra}, {Liebert}, {Lonsdale}, {Monet}, {Price}, {Seitzer}, {Jarrett}, {Kirkpatrick}, {Gizis}, {Howard}, {Evans}, {Fowler}, {Fullmer}, {Hurt}, {Light}, {Kopan}, {Marsh}, {McCallon}, {Tam}, {Van Dyk}, \& {Wheelock}}]{2mass}
{Skrutskie}, M.~F., {Cutri}, R.~M., {Stiening}, R., {et~al.} 2006, \aj, 131, 1163

\bibitem[{{Sollima} {et~al.}(2022){Sollima}, {Gratton}, {Lucatello}, \& {Carretta}}]{sollima22}
{Sollima}, A., {Gratton}, R., {Lucatello}, S., \& {Carretta}, E. 2022, \mnras, 512, 776

\bibitem[{{Tarricq} {et~al.}(2022){Tarricq}, {Soubiran}, {Casamiquela}, {Castro-Ginard}, {Olivares}, {Miret-Roig}, \& {Galli}}]{tarricq22}
{Tarricq}, Y., {Soubiran}, C., {Casamiquela}, L., {et~al.} 2022, \aap, 659, A59

\bibitem[{{Thomas} {et~al.}(2024){Thomas}, {Battaglia}, {Gran}, {Fern{\'a}ndez-Alvar}, {Tsantaki}, {Pancino}, {Hill}, {Kordopatis}, {Gallart}, {Turchi}, \& {Masseron}}]{thomas24}
{Thomas}, G.~F., {Battaglia}, G., {Gran}, F., {et~al.} 2024, \aap, 690, A54

\bibitem[{{VandenBerg} {et~al.}(2013){VandenBerg}, {Brogaard}, {Leaman}, \& {Casagrande}}]{vandenberg13}
{VandenBerg}, D.~A., {Brogaard}, K., {Leaman}, R., \& {Casagrande}, L. 2013, \apj, 775, 134

\bibitem[{{Vasiliev} \& {Baumgardt}(2021)}]{vasiliev21}
{Vasiliev}, E. \& {Baumgardt}, H. 2021, \mnras, 505, 5978

\bibitem[{{Webb} {et~al.}(2017){Webb}, {Vesperini}, {Dalessandro}, {Beccari}, {Ferraro}, \& {Lanzoni}}]{webb17}
{Webb}, J.~J., {Vesperini}, E., {Dalessandro}, E., {et~al.} 2017, \mnras, 471, 3845

\bibitem[{{Xiang} \& {Rix}(2022)}]{xiang22}
{Xiang}, M. \& {Rix}, H.-W. 2022, \nat, 603, 599

\end{thebibliography}

\begin{appendix}\label{appendix}

\section{Observing logs}
\begin{table*}[!h]
\centering \small
\caption{Observing log for the ACS sample.}\label{tab_acs}
\vspace{0.1 cm}
\vspace{0.1 cm}

{
\resizebox{\columnwidth}{!}{%
\begin{tabular}{lccccc}
\hline    
Name & $\alpha_{\rm 2000}$  &  $\delta_{\rm 2000}$  &  UT Date  &  Filter  &  Exposure Time  \\  
     &  (h m s)      &    ($^{\circ}$ $^\prime$ $^{\prime\prime}$)    &   &  & 
\\
\hline
\\
WHITING~1 & 02 02 57.0 & $-$03 15 10.0 & 2024-06-19  & F814W & 1\,$\times$\,30\,s, 2\,$\times$\,699\,s, 1\,$\times$\,567\,s\\
\vspace{0.1 cm}
          &            &             &             & F606W &  \\
KOPOSOV~1 & 11 59 18.4 & $+$12 15 35.8 & 2024-03-26  & F606W & 1\,$\times$\,30\,s, 2\,$\times$\,699\,s, 1\,$\times$\,567\,s\\
\vspace{0.1 cm}
        &              &             & 2024-05-22  & F814W & 1\,$\times$\,30\,s, 2\,$\times$\,699\,s, 1\,$\times$\,567\,s\\
KOPOSOV~2 & 07 58 17.0 & $+$26 15 19.4 & 2024-03-26  & F606W & 1\,$\times$\,30\,s, 2\,$\times$\,699\,s, 1\,$\times$\,575\,s\\
\vspace{0.1 cm}
        &              &             & 2024-05-22  & F814W & 1\,$\times$\,30\,s, 2\,$\times$\,699\,s, 1\,$\times$\,575\,s\\
MU\~{N}OZ~1 & 15 01 48.0 & $+$66 58 07.3 & 2025-03-15  & F606W, F814W & 1\,$\times$\,50\,s, 3\,$\times$\,766\,s\\
\vspace{0.1 cm}
        &              &             &  &  & \\
ESO-280-6 & 18 09 06.0 &$-$46 25 24.0 & 2025-04-08 & F606W, F814W & 1\,$\times$\,30\,s, 2\,$\times$\,699\,s, 1\,$\times$\,694\,s\\
\vspace{0.1 cm}
        &              &             &  &  & \\
Pal~8 & 18 41 29.9    &$-$19 49 33.1 & 2024-03-16  & F606W & 1\,$\times$\,30\,s, 2\,$\times$\,699\,s, 1\,$\times$\,570\,s\\
\vspace{0.1 cm}
        &              &             & 2024-03-24  & F814W & 1\,$\times$\,30\,s, 2\,$\times$\,699\,s, 1\,$\times$\,570s  \\
Pal~11 &  19 45 14.4  &$-$08 00 26.1 & 2024-03-14  & F606W &1\,$\times$\,30\,s, 2\,$\times$\,699\,s, 1\,$\times$\,566\,s\\
\vspace{0.1 cm}
        &              &             & 2024-03-17  & F814W &1\,$\times$\,30\,s, 2\,$\times$\,699\,s, 1\,$\times$\,566\,s\\
Laevens~3 &  21 06 54.3 &  $+$14 58 48.0 & 2025-04-16 & F606W &1\,$\times$\,30\,s, 2\,$\times$\,699\,s, 1\,$\times$\,567\,s\\
\vspace{0.1 cm}
        &              &             & 2025-04-19  & F814W &1\,$\times$\,30\,s, 2\,$\times$\,699\,s, 1\,$\times$\,567\,s\\
NGC~7492 &  23 08 26.7 &   $-$15 36 41.3 & 2024-05-16 & F606W & 1\,$\times$\,30\,s, 2\,$\times$\,699\,s, 1\,$\times$\,570\,s\\
\vspace{0.1 cm}
        &              &                 & 2024-05-12 & F814W & 1\,$\times$\,30\,s, 2\,$\times$\,699\,s, 1\,$\times$\,570\,s\\
NGC~6749 & 19 05 15.4 &   $+$01 53 59.9 & 2024-03-09 & F606W  & 1\,$\times$\,30\,s, 2\,$\times$\,699\,s, 1\,$\times$\,337\,s\\
\vspace{0.1 cm}
        &              &                & 2024-03-12 & F814W  & 1\,$\times$\,30\,s, 2\,$\times$\,699\,s, 1\,$\times$\,337\,s\\
FSR~1716 &  16 10 33.0 &  $-$53 44 12.0 & 2024-01-25 & F606W  & 1\,$\times$\,80\,s, 2\,$\times$\,699\,s, 1\,$\times$\,337\,s\\
\vspace{0.1 cm}
        &               &               & 2024-02-14 & F814W  & 1\,$\times$\,80\,s, 2\,$\times$\,699\,s, 1\,$\times$\,337\,s\\
RLGC~1 & 16 17 08.4     &  $-$44 35 38.6 & 2025-02-17 & F606W, F814W &1\,$\times$\,30\,s, 2\,$\times$\,699\,s, 1\,$\times$\,647\,s\\
\vspace{0.1 cm}
        &           &               &  &  & \\
Gran~2 & 17 11 33.6 & $-$24 50 56.4 & 2024-02-06 & F606W, F814W & 1\,$\times$\,30\,s, 2\,$\times$\,699\,s, 1\,$\times$\,570\,s\\
\vspace{0.1 cm}
        &           &               &  &  & \\
Gran~3 &  17 05 01.4 & $-$35 29 45.6 & 2025-02-13  &F606W, F814W  & 1\,$\times$\,50\,s, 2\,$\times$\,699\,s, 1\,$\times$\,560\,s\\
\vspace{0.1 cm}
        &           &               &  &  & \\
Gran~4 & 18 32 27.1 &  $-$23 06 50.4 & 2024-03-17 & F606W & 1\,$\times$\,30\,s, 2\,$\times$\,699\,s, 1\,$\times$\,570\,s\\
\vspace{0.1 cm}
        &             &               & 2024-03-21 & F814W & 1\,$\times$\,30\,s, 2\,$\times$\,699\,s, 1\,$\times$\,570\,s\\
Garro~1 & 14 09 00.0 & $-$65 37 12.0 & 2024-05-22 & F606W &1\,$\times$\,40\,s, 3\,$\times$\,699\,s\\
\vspace{0.1 cm}
        &           &             & 2025-02-21     & F814W & 1\,$\times$\,40\,s, 3\,$\times$\,699\,s\\
Gran~5 & 17 48 54.7 &  $-$24 10 12.0 & 2024-03-22 & F606W, F814W & 1\,$\times$\,30\,s, 2\,$\times$\,699\,s, 1\,$\times$\,337\,s\\
\vspace{0.1 cm}
&\\
ESO-452-11 &  16 39 25.5 & $-$28 23 52.1 & 2024-02-14 & F606W & 1\,$\times$\,80\,s, 2\,$\times$\,699\,s, 1\,$\times$\,337\,s\\
\vspace{0.1 cm}
        &                &               & 2024-01-29 & F814W & 1\,$\times$\,80\,s, 2\,$\times$\,699\,s, 1\,$\times$\,337\,s\\
Gran~1 & 17 58 36.6 &  $-$32 01 10.8 & 2024-03-21 & F606W, F814W  & 1\,$\times$\,80\,s, 2\,$\times$\,699\,s, 1\,$\times$\,337\,s\\
Pal~10 & 19 18 02.1 &  $+$18 34 17.9 & 2024-03-12 & F606W & 1\,$\times$\,30\,s, 2\,$\times$\,699\,s, 1\,$\times$\,337\,s\\
\vspace{0.1 cm}
        &          &              &  2024-03-16 & F814W & 1\,$\times$\,30\,s, 2\,$\times$\,699\,s, 1\,$\times$\,337\,s\\
BH~140 &  12 53 00.3 &  $-$67 10 28.0 & 2024-04-03 & F606W & 1\,$\times$\,30\,s, 2\,$\times$\,699\,s, 1\,$\times$\,337\,s\\
\vspace{0.1 cm}
        &           &              & 2024-05-24 & F814W & 1\,$\times$\,30\,s, 2\,$\times$\,699\,s, 1\,$\times$\,337\,s\\
Patchick~126 &  17 05 38.6 &  $-$47 20 32.0 & 2024-05-21 & F606W & 1\,$\times$\,50\,s, 2\,$\times$\,699\,s, 1\,$\times$\,337\,s\\
\vspace{0.1 cm}
        &                   &              & 2024-01-31 & F814W & 1\,$\times$\,50\,s, 2\,$\times$\,699\,s, 1\,$\times$\,337\,s\\
Bliss~1 &  11 50 02.6 & $-$41 46 19.2 & 2025-01-08 & F606W, F814W & 1\,$\times$\,30\,s, 2\,$\times$\,699\,s, 1\,$\times$\,647\,s\\
\vspace{0.1 cm}
&\\
Kim~3 &  13 22 45.2& $-$30 36 03.5 & 2024-05-19 & F606W, F814W & 1\,$\times$\,30\,s, 2\,$\times$\,699\,s, 1\,$\times$\,590\,s\\
\vspace{0.1 cm}
&\\
Segue~3 & 21 21 31.0 &   $+$19 07 02.0 & 2025-04-30 & F606W &1\,$\times$\,30\,s, 2\,$\times$\,699\,s, 1\,$\times$\,570\,s \\
\vspace{0.1 cm}
        &              &             & 2025-05-03  & F814W &1\,$\times$\,30\,s, 2\,$\times$\,699\,s, 1\,$\times$\,570\,s\\
PWM~2 & 17 58 39.0 &  $-$05 04 18.0 & 2025-03-20 & F606W & 1\,$\times$\,30\,s, 2\,$\times$\,699\,s, 1\,$\times$\,566\,s\\
\vspace{0.1 cm}
          &            &             & 2025-03-29 & F814W & 1\,$\times$\,30\,s, 2\,$\times$\,699\,s, 1\,$\times$\,566\,s \\
BH~176 & 15 39 07.4&$-$50 03 09.8 &  2024-03-09 & F606W, F814W & 1\,$\times$\,50\,s, 2\,$\times$\,699\,s, 1\,$\times$\,560\,s\\\\
\hline
\end{tabular}
}
}
\tablefoot{Exposure times and filters are the same for the parallel fields observed with the WFC3/UVIS camera. Target positions are taken from \citep[][2010 edition]{harris96}, or from \citet{baumgardt21} when missing in the former catalogue. The MGCS clusters without an observing date have not yet been imaged by the telescope at the time of the writing of this paper.}
\end{table*}

\begin{table*}[!htbp]
\centering\small
\caption{Observing log for the WFC3/IR sample.}\label{tab_wfc3ir}
\vspace{0.1 cm}
\vspace{0.1 cm}

{
\resizebox{\columnwidth}{!}{%
\begin{tabular}{lccccc}
\hline
Name & $\alpha_{\rm 2000}$  &  $\delta_{\rm 2000}$  &  UT Date  &  Filter  &  Exposure Time  \\  
     &  (h m s)      &    ($^{\circ}$ $^\prime$ $^{\prime\prime}$)    &   &  & 
\\
\hline
\\
\vspace{0.1 cm}
UKS~1 &  17 54 27.2 &  $-$24 08 43.0 & 2024-03-26  & F125W, F160W & 4\,$\times$\,300\,s, 4\,$\times$\,250\,s\\
\vspace{0.1 cm}
&\\
VVV-CL001 & 17 54 42.5 &  $-$24 00 53.0 & 2025-02-18 & F125W, F160W & 4\,$\times$\,300\,s, 4\,$\times$\,250\,s\\
\vspace{0.1 cm}
&\\
RLGC-02  &  18 45 28.2 & $-$05 11 33.3 & 2024-03-13 & F125W, F160W & 4\,$\times$\,300\,s, 4\,$\times$\,250\,s\\
\vspace{0.1 cm}
&\\
2MASS-GC01 &  18 08 21.8 &  $-$19 49 47.0 &2024-03-26 & F125W, F160W & 4\,$\times$\,300\,s, 4\,$\times$\,250\,s\\
\vspace{0.1 cm}
&\\
2MASS-GC02 & 18 09 36.5 & $-$20 46 44.0 & 2024-03-27 & F125W, F160W & 4\,$\times$\,300\,s, 4\,$\times$\,250\,s\\
\vspace{0.1 cm}
&\\
VVV-CL002 &  17 41 6.3 & $-$28 50 42.3 & 2024-03-20  & F125W, F160W & 4\,$\times$\,300\,s, 4\,$\times$\,250\,s\\
\vspace{0.1 cm}
&\\
Mercer~5 &  18 23 19.8 &  $-$13 40 07.1 & 2024-03-14 & F125W, F160W & 4\,$\times$\,300\,s, 4\,$\times$\,250\,s\\
&\\
\hline
\end{tabular}
}
}
%\tablefoot{The observations of the WFC3/IR sample are complete.}
\end{table*}

\newpage
\section{Differential reddening corrected CMDs}
\begin{figure*}
    \centering
    \includegraphics[width=0.7\textwidth]{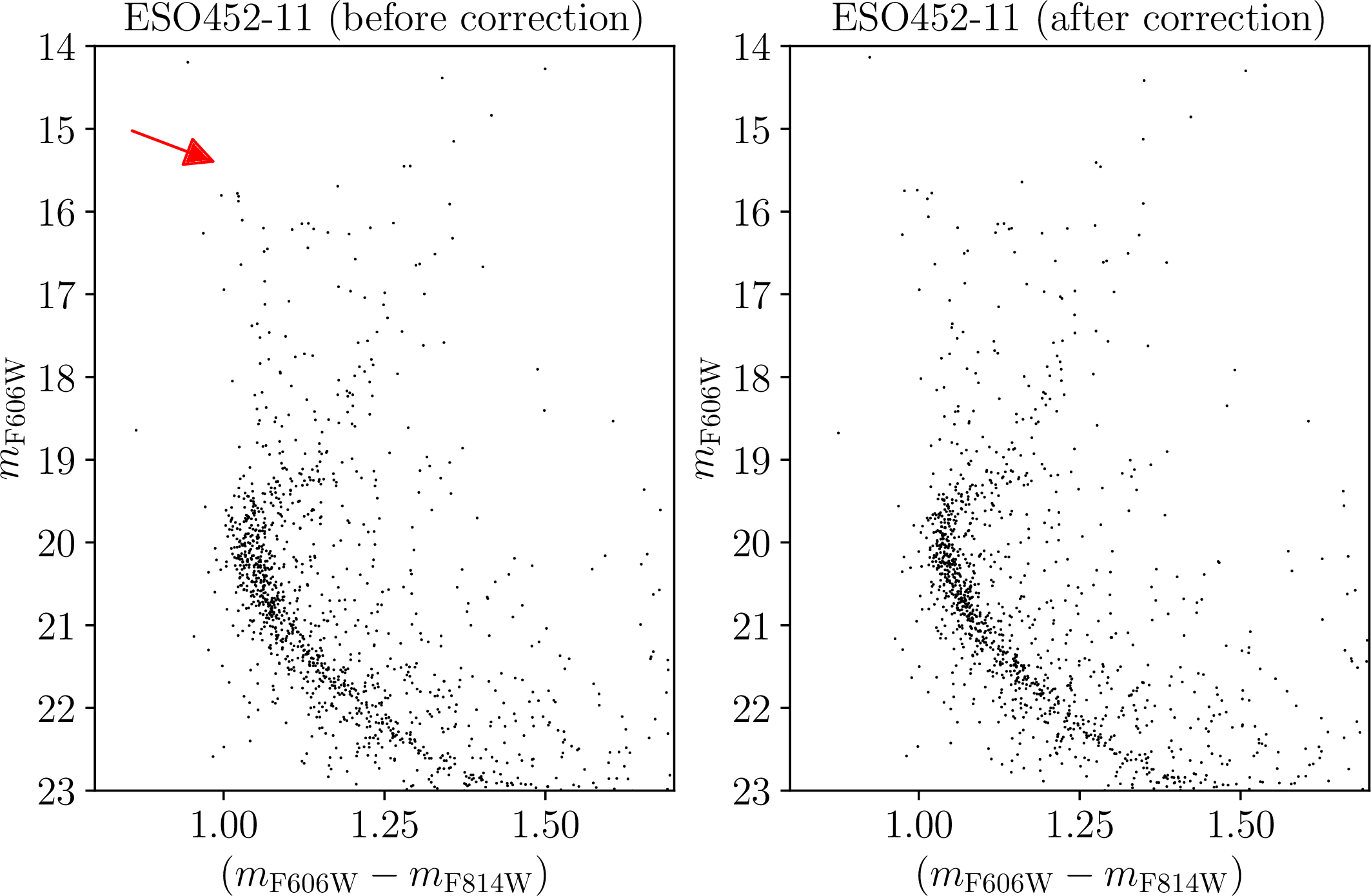}\vspace{1cm}
    \includegraphics[width=0.7\textwidth]{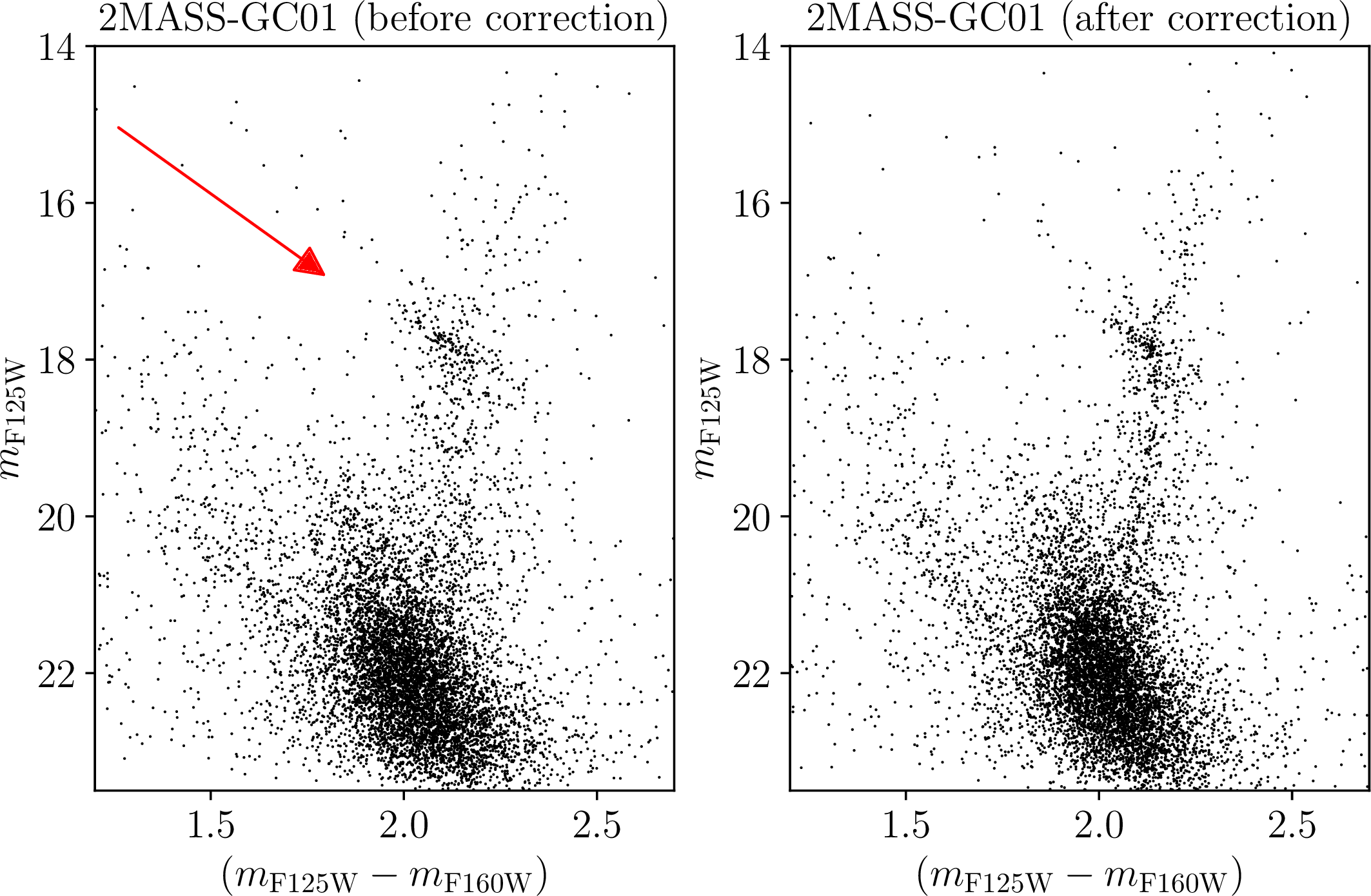}        
    \caption{\small Comparison between the CMDs of ESO452-11 (upper row) and 2MASS-GC01 (lower row), before (left-hand panels) and after (right-hand panels) applying our differential-reddening correction. The red arrows represent the reddening vector obtained using the $E(B-V)$ values from the literature (Sect.~\ref{sect:age}), the $A_\lambda$ assuming the \citet{cardelli89} law, and $R_V = 3.1$. The size of the arrows is reduced by a factor of three for clarity.} 
    \label{fig:cmddr}
\end{figure*}

Figure ~\ref{fig:cmddr} shows the CMDs of the two clusters studied here, before and after the differential reddening correction. The corresponding reddening maps are shown in Sect.~\ref{sec:results}.

\end{appendix}

\end{document}